%% file: main.tex
\documentclass[12pt]{article}
\usepackage{a4}
\usepackage{moreverb,relsize,url}

\usepackage{pslatex}

\usepackage{graphicx}
\usepackage[listings,skins]{tcolorbox}

\begin{document}

\title{Performance optimization and modeling of fine-grained irregular communication in UPC}

\author{J\'er\'emie Lagravi\`ere$^{1}$\and
Johannes Langguth$^{1}$ \and
Martina Prugger$^{2}$\footnote{The current affiliation of Martina Prugger
  is Vanderbilt University, 2201 West End Ave, Nashville, TN 37235, USA.} \and
Lukas Einkemmer$^{2}$ \and
Phuong H. Ha$^{3}$ \and
Xing Cai$^{1,4}$}

\date{
{\small $^{1}$Simula Research Laboratory, P.O.~Box 134, NO-1325 Lysaker, Norway}\\
{\small $^{2}$University of Innsbruck, Technikerstra{\ss}e 13,  A-6020 Innsbruck,
Austria}\\
{\small $^{3}$The Arctic University of Norway, NO-9037 Troms{\o}, Norway}\\
{\small $^{4}$University of Oslo, NO-0316 Oslo, Norway}\\
\vspace{.5cm}
{\scriptsize Correspondence should be addressed to Xing Cai; xingca@simula.no}
}
\maketitle

\begin{abstract}

The UPC programming language offers parallelism via logically partitioned shared
memory, which typically spans physically disjoint memory sub-systems. 
One convenient feature of UPC is
its ability to automatically execute between-thread data movement,
such that the entire content of a shared data array appears to be freely
accessible by all the threads. The
programmer friendliness, however, can come at the cost of substantial
performance penalties. This is especially true when indirectly indexing
the elements of a shared array, for which the induced between-thread data
communication can be irregular and have a fine-grained pattern.
In this paper we study performance enhancement strategies specifically targeting
such fine-grained irregular communication in UPC.
%that is associated with indirect indexing of shared data arrays in UPC. 
Starting from explicit thread privatization,
%the standard approach of casting shared array pointers to private local pointers, 
continuing with block-wise communication, and arriving at message condensing and
consolidation, we obtained considerable performance improvement of UPC
programs that originally require fine-grained irregular communication.
Besides the performance enhancement strategies, the main contribution of the present paper is to propose performance models for the different scenarios, 
in form of quantifiable formulas that hinge on the actual volumes of
various data movements
plus a small number of easily obtainable hardware characteristic parameters.
These performance models help to verify the
enhancements obtained, while also providing insightful predictions of
similar parallel implementations, not limited to UPC, that also involve between-thread or
between-process irregular communication. As a further validation,
we also apply our performance modeling methodology and hardware
characteristic parameters to an existing UPC code for solving a 2D
heat equation on a uniform mesh.


\end{abstract}

\input{Sec_Motivation.tex}
\input{Sec_Shared_array.tex}

\input{Sec_SpMV.tex}
\input{Sec_Enhancements.tex}
\input{Sec_PerfModels.tex}
\input{Sec_Experiments.tex}

\input{Sec_Related_work.tex}
\input{Sec_Heat2D.tex}

\input{Sec_Conclusion.tex}

\vspace{.8cm}
{\bf Conflicts of interest}:
The authors declare that there is no conflict of interest regarding
the publication of this paper.

{\bf Acknowledgements}:
This work was performed on hardware resources provided by UNINETT
Sigma2 -- the National Infrastructure for High Performance Computing
and Data Storage in Norway, via Project NN2849K. 
The work was supported by the European Union's Horizon 2020 research and innovation programme under grant agreement No.~671657, the European Union Seventh Framework Programme (grant No.~611183) and the Research Council of
Norway (grants No.~231746/F20, No.~214113/F20 \& No.~251186/F20).
Martina Prugger was supported by the VSC ResearchCenter funded by the Austrian Federal Ministry of Science, Research and Economy (bmwfw) and by the Doctoral Programme Computational Interdisciplinary Modelling (DK CIM) at the University of Innsbruck.

\bibliographystyle{plain}
\bibliography{UPC_references}

\end{document}

%% file: Sec_Motivation.tex
\section{Motivation}    %%%{Introduction}

Good programmer productivity and high computational performance are usually two
conflicting goals in the context of developing parallel code for scientific computations.
{\em Partitioned global address space}
(PGAS)~\cite{1263470,Almasi2011,PGAS_survey,PGAS_site},
however, is a parallel programming model that aims to achieve both goals at the same time.
The fundamental mechanism of PGAS is a 
global address space that is conceptually shared among concurrent processes
that jointly execute a parallel program. 
Data exchange between the processes is carried out by a low-level network layer
``under the hood'' without explicit involvement from the programmer,
thus providing good productivity.
The shared global address space is logically partitioned 
such that each partition has affinity to a designated owner process. This awareness of %strategy of promoting awareness of
data locality is essential for achieving good performance of parallel
programs written in the PGAS model, because the globally shared
address space may actually encompass many physically distributed memory sub-systems.
%incurring high overhead for non-private memory operations
%especially when programmed inappropriately.

{\em Unified Parallel C} (UPC)~\cite{UPC_book,UPC_specification} is an extension of the C %programming
language and provides the PGAS parallel programming model.
The concurrent execution processes of UPC are termed as {\em
  threads}, which execute a UPC program in the style of {
  single-program-multiple-data}. The data variables of each thread
are of two types: {\em private} and {\em shared}.
Variables of the second type, accessible by 
 all the threads, are found in the globally shared address space.
In particular, shared data arrays of UPC provide programmer friendliness because
any thread can use a global index to access an arbitrary array element.
If the accessing thread does not own the target
element, between-thread communication will be carried out automatically. 

Another parallelization-simplifying feature of UPC is that shared arrays allow a
straightforward distribution of data ownership among the threads.
%, see Section~\ref{sec:two}.
However, the simple data distribution scheme adopted for shared arrays may bring
disadvantages. First, balancing the computational work among the
threads can be more challenging than other parallel programming models that
allow an uneven  (static or
dynamic) distribution of array elements to account for the possibly
inhomogeneous cost per element. Second, for a UPC program where
between-thread memory operations are inevitable, which is true for
most scientific applications, the only mechanism for a programmer to
indirectly control the impact of %overhead due to
remote memory traffic is to tune the block size constant of shared arrays. Third,
all non-private memory operations (i.e.~between threads)
are considered in UPC to be of one type. There is
no way for UPC to distinguish intra-compute-node memory operations
(between threads running on the same hardware node)
from their inter-node counterparts.
% The intra-node operations are between threads running on
% %(different sockets of) 
% the same hardware node that has a
% physically shared NUMA memory, thus considerably faster than
% the inter-node remote memory operations.
The latter require explicitly
transferring data over some interconnect between the nodes, 
which is considerably more costly.

In this paper, we will closely investigate the second and third disadvantages
mentioned above. This will be done in the context of fine-grained remote memory
operations that have irregular thread-to-thread communication
patterns. They can arise from irregular and indirectly indexed
accesses to the elements of shared arrays in UPC.
Our objectives are two-fold. First, we want to study the impact of
several performance-enhancing techniques in UPC programming:
privatizing for-loop iterations among threads, 
explicitly casting pointers-to-shared to pointers-to-local, 
adopting bulk memory transfers instead of individual 
remote-memory accesses,
and message condensing with consolidation. Second, and more
importantly, we will propose performance models for a representative
example of scientific computations that induce fine-grained and
irregular UPC remote memory operations.
Based on a simple philosophy of quantifying the occurrences and volumes
of two categories of inter-thread memory traffic: local inter-thread traffic
(within a compute node) and remote inter-thread traffic (between nodes),
 these performance models only need
a small number of hardware characteristic parameters to provide
a % theoretical and yet
realistic performance prediction. This helps to
understand and further tune the obtainable performance
on an existing hardware system, while giving insightful
predictions of the achievable scalability on upcoming new platforms.

We will focus on a specific category of matrix-vector multiplication operations
where the involved matrix is sparse and has a constant number of
nonzero values per row. Such a computational kernel appears in many branches of
computational science. % For instance, it may arise from discretizing a partial
% differential equation on an unstructured computational mesh.
% by a cell-centered finite volume method. 
 A performance challenge
is that the nonzero values per matrix row can spread irregularly with respect
to the columns. Consequently, any computer implementation will involve
 irregular and indirectly indexed accesses to numerical arrays. The
resulting fine-grained irregular data accesses need
to be handled with care, particularly in parallel implementations.
We thus choose this computational kernel
as a concrete case of fine-grained irregular communication that can occur in UPC code.
We want to demonstrate that proper code transformations can be applied to a
naive, programmer-friendly but inefficient UPC implementation, for obtaining
considerable enhancements of the computing speed.
Moreover, the obtained performance enhancements can be backed up by
conceptually simple performance models.

The remainder of this paper is organized as follows.
Section~\ref{sec:two} explains the basic mechanism of shared arrays, the cornerstone of
UPC programming. Then Section~\ref{sec:spmv} gives a numerical description of
our target computational problem and a naive UPC
implementation. Thereafter, Section~\ref{sec:transforms} shows in detail three
programming strategies that transform the naive implementation for
increasingly better performance. In Section~\ref{sec:perf_model}, three
performance models are developed to match with the three code
transformations. Section~\ref{sec:experiments} presents an extensive set of
numerical experiments and time measurements for
both showing the impact of the code transformations and verifying the
performance models developed. 
Relevant related work is reviewed in
Section~\ref{sec:related_work}, whereas Section~\ref{sec:heat2D} shows
how our performance modeling methodology can easily be extended to
simpler 2D calculations on a uniform mesh,
before Section~\ref{sec:conclusion} concludes
the entire paper with additional comments.

%% file: Sec_Shared_array.tex
\section{Shared arrays of UPC}
\label{sec:two}

%In UPC, shared scalar variables all have affinity to thread 0.
% Although elements of a
% shared array can have exclusive affinity to a single thread (i.e.,~all
% the elements are allocated in the local memory of the
% single owner thread),

Shared data arrays, whose elements are freely accessible by all the threads, 
typically constitute the main data structure of a UPC program.
They thus deserve a separate introduction in this section.
The most common scenario is that the elements of a 
shared array have an evenly distributed thread affinity.
%(instead of having affinity to a single thread). 
This gives a straightforward approach to data partitioning while
providing a user-controllable mechanism of data locality.
The standard  {\tt upc\_all\_alloc} function (see
e.g.~\cite{UPC_specification}), to be used by all the
UPC implementations in this paper for array allocation, has the following syntax:
\begin{quote}
\begin{verbatim}
shared void *upc_all_alloc(size_t nblks, size_t nbytes);
\end{verbatim}
\end{quote}
Note that {\tt shared} is a specific type qualifier.
Also, {\tt upc\_all\_alloc} needs
 to be {\em collectively} called by all threads to allocate a shared array.
Upon return, a private pointer to the allocated shared array, 
or a private {\em pointer-to-shared} in the more rigorous UPC
terminology, becomes available on each thread.
The allocated shared array consists of {\tt
  nblks} blocks in total, whose affinity is distributed evenly among the
threads in a {\em cyclic} order. 
The value of {\tt nbytes} is the number of bytes occupied per block,
which translates the block size, i.e., number of elements per block,
as {\tt nbytes/sizeof}(one element).
The blocks that have affinity to the same owner
thread are physically allocated {\em contiguously} in the owner thread's local
memory.
%
% The affinity distribution scheme is simple, following a blocked cyclic
%  partitioning, where the block size of %the affinity distribution for 
% a shared array remains constant throughout its lifetime.

This data ownership distribution scheme, which in
many cases determines an associated work partitioning, has the
advantage of a simple mapping between a global element index and the
owner thread ID. It can be described as follows:
\begin{eqnarray}
\mbox{\tt owner\_thread\_id} = 
\Bigl\lfloor\frac{\mbox{\tt global\_index}}{\mbox{\tt  block\_size}}\Bigr\rfloor
  \; \mathrm{modulo} \; \mbox{\tt THREADS} \, ,
\label{affinity_mapping}
\end{eqnarray}
where {\tt THREADS} is a built-in variable of UPC that stores the
total number of threads participating in a parallel execution. 
The value of {\tt THREADS} is fixed at either compile time or run-time.

Accessing the elements of a shared array by their global indices,
although programmer friendly, can potentially incur considerable overhead. 
This is because a pointer-to-shared has three fields: the
owner thread ID, the phase (i.e.~the element offset within the affinity
block) and the corresponding local memory address, see~\cite{UPC_book}.
The standard {\tt
  upc\_threadof(shared void *ptr)} function of UPC returns the owner thread ID of an
element that is pointed to by the pointer-to-shared {\tt ptr}.
Every access through a pointer-to-shared
requires updating the three fields and thus always incurs overhead.
Moreover, if the accessing thread is
different from the owner thread, a behind-the-scene data transfer
between the two threads has to be carried out. For
indirectly indexed accesses of the elements in a shared array, 
%such as {\tt x[J[i*r\_nz+j]]},
a compiler cannot batch the individual
between-thread data exchanges for the purpose of message aggregation
(such as described in~\cite{Chen2005,Chen_PhD2007}).
The individual between-thread data exchanges are thus particularly costly
when the accessing thread runs on a different compute node than the owner thread.

%% file: Sec_SpMV.tex
\section{SpMV and a naive UPC implementation}
\label{sec:spmv}

This section is devoted to explaining the target computational kernel
of this paper and presenting a naive UPC implementation.

\subsection{Definition of sparse matrix-vector multiplication}
\label{sec:3.1}
 Mathematically, a general matrix-vector multiplication is compactly
denoted by $y=M x$.
Without loss of generality we assume that the matrix $M$ is square,
having $n$ rows and $n$ columns. 
The input vector $x$ and the result vector $y$ are both
of length $n$.
Then, the general formula for computing element number $i$ of the
result vector $y$ is as follows (using zero-based indices):
\begin{equation}
y(i) = \sum_{0\le j< n} M(i,j)x(j).
\label{eqn:MV}
\end{equation}

If most of the $M(i,j)$ values are zero, $M$ is called a {\em sparse}
matrix. In this case the
above formula becomes unnecessarily expensive from a computational
point of view.
A more economic formula for computing $y(i)$ in a sparse matrix-vector
multiplicatin (SpMV) is thus
\begin{equation}
y(i) = \sum_{M(i,j)\neq 0} M(i,j)x(j),
\label{eqn:SpMV}
\end{equation}
which only involves the nonzero values of matrix $M$ on each row.
Moreover, it is memory-wise unnecessarily expensive to store all
the $n^2$ values of a sparse matrix, because only the nonzero values
are used. This prompts the adoption of various compact storage
formats for sparse matrices, 
such as the coordinate format (COO), compressed sparse row
format (CSR), compressed sparse column format (CSC), and the EllPack
format~\cite{Grimes1979}.

In particular, for sparse matrices that have a fixed number of nonzero values per
row, it is customary to use the EllPack storage format, which
conceptually uses two
2D tables. Both tables are of the same size, having $n$ rows and
the number of columns equaling the fixed number of nonzeros per row.
The first table stores all the nonzero values of the sparse matrix,
whereas the second table stores the corresponding column indices of
the nonzeros. 
Moreover, if we assume that all the values on the
main diagonal of a sparse matrix $M$ are nonzero, which is true for
most scientific applications, it is beneficial to split $M$ as
\begin{equation}
M=D+A, \label{eqn:SpMV2}
\end{equation}
where $D$ is the main diagonal of $M$, and $A$ contains the off-diagonal part
of $M$. Then, a modified EllPack storage format can employ a 1D array of
length $n$ to store the entire main diagonal $D$.
There is no need to store the column indices of these nonzero
diagonal values, because their column indices equal the row indices by
definition. Suppose $r_{nz}$ now denotes the fixed number of nonzero
off-diagonal values per row.
For storing the nonzero values in the off-diagonal part $A$, it is
customary to use two 1D arrays
 both of length $n\cdot r_{nz}$ (instead of two $n\times r_{nz}$ 2D tables),
one stores the
nonzero off-diagonal values consecutively row by row, whereas the other stores the
corresponding integer column indices.

Following such a modified EllPack storage format, a straightforward
sequential C implementation of SpMV is shown in Listing~\ref{code-seq-plain},
where the integer array {\tt J} contains the column indices of all nonzero
off-diagonal values.

\begin{lstlisting}[caption=A straightforward sequential implementation
  of SpMV using a modified EllPack storage format,
%  of SpMV $v_{\mathrm new}=(D+A)v$ using an EllPack based storage format, 
label=code-seq-plain]
for (int i=0; i<n; i++) {
  double tmp = 0.0;
  for (int j=0; j<r_nz; j++)
    tmp += A[i*r_nz+j]*x[J[i*r_nz+j]];
  y[i] = D[i]*x[i] + tmp;
}
\end{lstlisting}

The sparsity pattern of the $M$ matrix, i.e., where its nonzeros are located,
% which values from the input
% vector $x$ are used to compute each $y(i)$ value of the result vector,
is described by the array {\tt J} of column indices. The actual
pattern is matrix dependent and irregular in general, meaning that
% we do know that each $x(i)$ value partipates in the computation of exactly
% $r_{nz}+1$ values of the $y$ vector, including the $y(i)$
% value. This is due to the assumption of having
% a const number of nonzeros per row in the $M$ matrix.
each $x(i)$ value is irregularly used multiple times in computing
 several values in the result vector $y$. This has an important bearing on the achievable
performance of a typical computer implementation, because the actual
sparsity pattern of the $M$ matrix may affect the level of data reuse in
the different caches of a computer's memory system. 
Additionally, for the case of parallel
computing, some of the values in the $x$ vector have to be shared
between processes (or threads). The irregular data reuse in the
$x$ vector will thus imply an irregular data sharing pattern. 
The resulting communication overhead is determined by the
number of process pairs that need to share some values of
the $x$ vector, as well as the amount of shared data between each pair.
The impact of these issues on different UPC implementations of
SpMV will be the main subject of study in this paper. In the following, we
first present a naive UPC implementation, whereas code transformation
strategies that aim to improve the performance will be discussed in Section~\ref{sec:transforms}.

\subsection{A naive UPC implementation}
\label{sec:3.2}

The user-friendliness of UPC allows for an equally compact and almost
identical implementation
of the SpMV computational kernel (starting from the line of {\tt
  upc\_forall} in Listing~\ref{code-UPC-naive}) as the
straightforward C implementation in Listing~\ref{code-seq-plain}.
 An immediate advantage is that
parallelization is automatically enabled through using the {\tt
  upc\_forall} construct of UPC, which deterministically divides the iterations of
a for-loop among the threads. The five
involved data arrays, which are all allocated by {\tt upc\_all\_alloc}
 as shared arrays, are evenly
distributed among the UPC threads in a block-cyclic order. More
specifically, the arrays {\tt x}, {\tt y} and {\tt D} adopt a programmer-prescribed
integer value, {\tt BLOCKSIZE}, as their block size associated with
the affinity distribution. The arrays {\tt A} and {\tt J}, both of
length $n\cdot r_{nz}$, use {\tt r\_nz*BLOCKSIZE} as their block
size. This gives a consistent thread-wise data distribution for
the five shared data arrays.

\begin{lstlisting}[caption=A naive UPC implementation
  of SpMV using a modified EllPack storage format,
label=code-UPC-naive]
/* Total number of blocks in every shared array */
int nblks = n/BLOCKSIZE + (n%BLOCKSIZE)?1:0;
/* Allocation of five shared arrays */
shared [BLOCKSIZE] double *x = upc_all_alloc (nblks, BLOCKSIZE*sizeof(double));
shared [BLOCKSIZE] double *y = upc_all_alloc (nblks, BLOCKSIZE*sizeof(double));
shared [BLOCKSIZE] double *D = upc_all_alloc (nblks, BLOCKSIZE*sizeof(double));
shared [r_nz*BLOCKSIZE] double *A = upc_all_alloc (nblks, r_nz*BLOCKSIZE*sizeof(double));
shared [r_nz*BLOCKSIZE] int *J = upc_all_alloc (nblks, r_nz*BLOCKSIZE*sizeof(int));
// ...
/* Computation of SpMV involving all threads */
upc_forall (int i=0; i<n; i++; &y[i]) {
  double tmp = 0.0;
  for (int j=0; j<r_nz; j++)
    tmp += A[i*r_nz+j]*x[J[i*r_nz+j]];
  y[i] = D[i]*x[i] + tmp;
}
\end{lstlisting}

The UPC implementation of SpMV shown in Listing~\ref{code-UPC-naive}
is clean and easy to code. The parallelization details,
i.e., data distribution and work partitioning,
are an inherent part of the language definition of UPC. 
Since the number of nonzeros per matrix row is assumed to be fixed, the adopted
thread-wise data distribution for the shared {\tt D}, {\tt A}, {\tt J} and {\tt y}
arrays is perfect, in that each thread will only access its owned blocks of
these arrays. For the shared array {\tt x}, whose values are indirectly
accessed via the column index array {\tt J},
underlying data transfers between the threads are inevitable in general.

As will be detailed later, the irregular
column positions of the nonzero values (stored in the array {\tt J})
 will cause fine-grained and
irregular between-thread data exchanges associated with the shared
array {\tt x} in the straightforward UPC
implementation shown in Listing~\ref{code-UPC-naive}.
Tuning the value of {\tt BLOCKSIZE} can change
the pattern and volume of between-thread communication. However, to ensure good
performance, proper code transformations of such a naive UPC
implementation are necessary.

% When the number of rows in the sparse matrix is very large, it is customary to 
% parallelize the entire computation. UPC can provide a concise and
% programmer-friendly implementation of SpMV where fine-grained
% irregular communication automatically happens between physically
% disjoint memory spaces

%% file: Sec_Enhancements.tex
\section{Strategies of performance enhancement}
\label{sec:transforms}

This section studies three programming strategies that can
be applied to transforming the naive UPC implementation. The main purpose
is to reduce the impact of implicit between-thread data exchanges that
are caused by irregular and indirectly indexed accesses to the shared
array {\tt x}. At the same time, we also want to eliminate some of the
other types of avoidable overhead associated with UPC programming.

\subsection{Explicit thread privatization}
\label{sec:4.1}

Some of the programmer-friendly features of UPC are accompanied with performance
penalties. Relating to the naive UPC implementation of SpMV in
Listing~\ref{code-UPC-naive},  these concern automatically
dividing the iterations of a for-loop among threads by {\tt upc\_forall} and
allowing any thread to access any element of a shared array.
See Section~\ref{sec:two} about the latter.

The {\tt upc\_forall} construct of UPC is a collective operation. In the
example of {\tt upc\_forall (i=0; i<n; i++; \&y[i])} used in
Listing~\ref{code-UPC-naive}, all the threads go
through the {\em entire} for-loop and check the affinity of each iteration,
by comparing whether {\tt upc\_threadof(\&y[i])} equals the built-in
{\tt MYTHREAD} value that is unique per thread.
Although only iterations having an affinity equaling {\tt MYTHREAD}
are executed by a thread, it is not difficult to see
the overhead due to excessive looping and calls to the standard {\tt
  upc\_threadof} function behind the scene.

% Calculations of form (\ref{affinity_mapping}) are also involved when
% a thread accesses an element of a shared array. 
% Addtional calculations are also needed to find the local memory
% location where the target element lies in its owner thread.
% If the accessing thread is different from the owner thread,
% between-thread data traffic will be generated. For the case of {\tt
%   x[J[i*r\_nz+j]]}, due to indirect indexing through the {\tt J}
% array, a compiler will normally not be able to aggregate the individual
% between-thread data exchanges. This makes every indirectly-indexed 
% access to a non-owner
% element of a shared array particularly costly.

To get rid of the unnecessary overhead associated with {\tt upc\_forall},
we can let each thread work directly with the loop iterations
that have the matching affinity. Note that
the affinity distribution of the {\tt i}-indexed loop iterations can be easily determined using the value of {\tt BLOCKSIZE}. %follows that of the shared {\tt y} array
Such an explicit thread-privatization of the loop iterations also
opens up another opportunity for performance enhancement. Namely, all the
globally indexed accesses to the shared arrays {\tt y}, {\tt A}, {\tt
  J} and {\tt D} (except {\tt x}) can be replaced by more efficient 
accesses through private pointers and local indices. This is achievable 
by using the well-known technique of casting
pointers-to-shared to pointers-to-local, see e.g.~\cite{Zheng2010}.
Following these two steps that can be loosely
characterized as {\em explicit thread privatization}, the naive UPC
implementation can be transformed as shown in Listing~\ref{code-UPC-v1}.

\begin{lstlisting}[caption=An improved UPC implementation
  of SpMV by explicit thread privatization,
label=code-UPC-v1]
/* Allocation of the five shared arrays x, y, D, A, J as in the naive implementation */
// ...
/* Instead of upc_forall, each thread directly handles its designated blocks */
int mythread_nblks = nblks/THREADS+(MYTHREAD<(nblks%THREADS)?1:0);
for (int mb=0; mb<mythread_nblks; mb++) {
  int offset = (mb*THREADS+MYTHREAD)*BLOCKSIZE;
  /* casting shared pointers to local pointers */
  double *loc_y = (double*) (y+offset);
  double *loc_D = (double*) (D+offset);
  double *loc_A = (double*) (A+offset*r_nz);
  int *loc_J = (int*) (J+offset*r_nz);
  /* computation per block */
  for (int k=0; k<min(BLOCKSIZE,n-offset); k++) {
    double tmp = 0.0;
    for (int j=0; j<r_nz; j++)
      tmp += loc_A[k*r_nz+j]*x[loc_J[k*r_nz+j]];
    loc_y[k] = loc_D[k]*x[offset+k] + tmp;
  }
}
\end{lstlisting}

It can be observed in Listing~\ref{code-UPC-v1} that each thread now only
traverses its designated rows of the sparse matrix. The computational work per
thread is executed by going through its owned blocks in the shared
arrays {\tt
  y}, {\tt D}, {\tt A} and {\tt J}, for which each thread is
guaranteed to never touch blocks owned by the other threads. Pointers to these four
shared arrays are cast to their local counterparts {\tt
  loc\_y}, {\tt loc\_D}, {\tt loc\_A} and {\tt loc\_J}, since array accesses through
private pointers and local indices are the most efficient. On the
other hand, casting pointer-to-shared {\tt x} cannot be done, because the indirect
accesses of form {\tt x[loc\_J[k*r\_nz+j]]} may lead to situations where the
accessing thread is different from the owner thread.
%% an element of x will be used by multiple rows of M, possibly
%% accessed by several non-owner threads
Compared with the naive UPC implementation in
Listing~\ref{code-UPC-naive}, the transformed version after explicit thread
privatization will have a much better performance. However, further
performance improvement can be obtained by also ``privatizing'' the
global accesses to the shared array {\tt x}. This can be achieved by
two code transformations,
of different programming complexities and performance
gains, which are to be detailed below.

\subsection{Block-wise data transfer between threads}
\label{sec:4.2}

Although Listing~\ref{code-UPC-v1}
improves the naive UPC implementation by the approaches of explicit
thread privatization, each thread still indirectly accesses the
elements of the shared array
{\tt  x} through global indices that are stored in
the array {\tt J} (now cast to pointer-to-local {\tt loc\_J} per block).
% The indirectly indexed accesses can be irregular due to the
% assumed irregular sparsity pattern of matrix $M$.
When an indirectly indexed {\tt x[loc\_J[k*r\_nz+j]]} value has
affinity with {\tt MYTHREAD}, the overhead only concerns updating the
three fields of a pointer-to-shared. If {\tt MYTHREAD} is different from the
owner thread, however, a behind-the-scene data transfer will be executed in
addition. Moreover, these between-thread data transfers will happen one by one,
because a typical compiler is unable to batch the individual transfers.
The extra overhead is particularly high if the owner and accessing threads
reside on different compute nodes.
To avoid the potentially high overhead associated with 
{\tt x[loc\_J[k*r\_nz+j]]}, we can create a private
copy of {\tt x} on each thread and transfer the needed blocks from
{\tt x} to the private copy before carrying out the SpMV. The resulting
UPC implementation is shown in Listing~\ref{code-UPC-v2}.

\begin{lstlisting}[caption=An improved UPC implementation
  of SpMV by block-wise communication,
label=code-UPC-v2]
/* Allocation of the five shared arrays x, y, D, A, J as in the naive implementation */
// ...
/* Allocation of an additional private x array per thread */
double *mythread_x_copy = (double*) malloc(n*sizeof(double));
/* Prep-work: check for each block of x whether it has values needed by MYTHREAD; make a private boolean array 'block_is_needed' of length nblks */
// ....
/* Transport the needed blocks of x into mythread_x_copy */
for (int b=0; b<nblks; b++)
   if (block_is_needed[b])
     upc_memget(&mythread_x_copy[b*BLOCKSIZE],&x[b*BLOCKSIZE],
                min(BLOCKSIZE,n-b*BLOCKSIZE)*sizeof(double));
/* SpMV: each thread only goes through its designated blocks */
int mythread_nblks = nblks/THREADS+(MYTHREAD<(nblks%THREADS)?1:0);
for (int mb=0; mb<mythread_nblks; mb++) {
  int offset = (mb*THREADS+MYTHREAD)*BLOCKSIZE;
  /* casting shared pointers to local pointers */
  double *loc_y = (double*) (y+offset);
  double *loc_D = (double*) (D+offset);
  double *loc_A = (double*) (A+offset*r_nz);
  int *loc_J = (int*) (J+offset*r_nz);
  /* computation per block */
  for (int k=0; k<min(BLOCKSIZE,n-offset); k++) {
    double tmp = 0.0;
    for (int j=0; j<r_nz; j++)
      tmp += loc_A[k*r_nz+j]*mythread_x_copy[loc_J[k*r_nz+j]];
    loc_y[k] = loc_D[k]*mythread_x_copy[offset+k] + tmp;
  }
}
\end{lstlisting}

In Listing~\ref{code-UPC-v2} we have used
 the one-sided communication function {\tt upc\_memget}
 of UPC to transfer all the needed
blocks, one by one, from the shared array {\tt x} into a thread-private local copy
named {\tt mythread\_x\_copy}. 
The syntax of  {\tt upc\_memget} is as follows:
\begin{quote}
{\tt void upc\_memget(void *dst, shared const void *src, size\_t n);}
\end{quote}
We have thus completely avoided 
accessing values of the shared array {\tt x}. However, there are a few
``prices'' paid on the way. First, each thread needs to
allocate its own {\tt mythread\_x\_copy} array of length $n$. 
This obviously increases the total memory usage. Second,
 the actual computation of SpMV on each thread
needs to be preceded by transporting all the needed blocks
from the shared array {\tt x} into the corresponding places of
{\tt mythread\_x\_copy}. Specifically, a needed
block from {\tt x} is defined as having at least one {\tt
  x[loc\_J[k*r\_nz+j]]} value that will
participate in calculating the designated elements in {\tt y} on each
thread (with {\tt MYTHREAD} as its unique ID).
We note that each needed block is transported in its {\em entirety}, independent of
the actual number of {\tt x} values needed in that block. This
also applies to the blocks of {\tt x} that are owned by
{\tt MYTHREAD}. The whole procedure of transporting the needed blocks of
{\tt x}, implemented as the for-loop indexed by {\tt b} in
Listing~\ref{code-UPC-v2}, will result in time usage
overhead. Nevertheless, this additional time usage is often
compensated by avoiding the individual accesses to the shared array {\tt
  x}.  Third, to identify whether a block of {\tt x} is needed
by {\tt MYTHREAD} requires pre-screening the designated blocks of the array {\tt
  J} (not shown in Listing~\ref{code-UPC-v2}). 
This is typically considered a negligible ``one-time'' cost if
the same sparse matrix, or the same sparsity pattern shared among
several sparse matrices, is repeatedly used in many SpMV operations later.

\subsection{Message condensing and consolidation}
\label{sec:4.3}

One shortcoming of the transformed UPC code shown in Listing~\ref{code-UPC-v2} is
that each needed block from {\tt x} is transported in its entirety. This
will lead to unreasonably large messages, when only a
small number of values in a block of {\tt x} are needed by {\tt
  MYTHREAD}.
Also, several messages may be transported
(instead of one consolidated message) between a pair of threads, where
each message has a rigid length of {\tt BLOCKSIZE}.
%except when the needed block is the last segment of {\tt x}.
To condense and consolidate the messages, we can carry out a different
code transformation as follows.

\subsubsection{Preparation step}
Each thread checks, in a ``one-time'' preparation step,
which of its owned {\tt x} values will be needed by the other
threads. We also ensure that only one message is exchanged between each pair of
communicating threads. The length of a message from thread {\tt T1}
to thread {\tt T2} equals the number of {\em unique} values in the {\tt x}
blocks owned by {\tt T1} that are needed by {\tt T2}.
All the between-thread messages are thus condensed and consolidated.
After this preparation step, the
following private arrays are created on each thread:
\begin{quote}
{\tt 
int *mythread\_num\_send\_values, *mythread\_num\_recv\_values;\\
int **mythread\_send\_value\_list, **mythread\_recv\_value\_list;\\
double **mythread\_send\_buffers;
}
\end{quote}
All the above private arrays have length {\tt THREADS} (in the leading direction).
If {\tt mythread\_num\_send\_values[T]}$>0$, it means that {\tt
  MYTHREAD} needs to pack an outgoing message of this length for thread {\tt T} as
the receiver. 
Correspondingly, 
 {\tt mythread\_send\_value\_list[T]} points to a list of local
indices relative to a pointer-to-local, which is cast from {\tt
  \&x[MYTHREAD*BLOCKSIZE]}, so that the respective needed {\tt x} values can
be efficiently extracted and packed together as the outgoing message {\tt
  mythread\_send\_buffers[T]}  toward thread {\tt T}.
The one-sided communication command {\tt upc\_memput}, which is of the following
syntax:
\begin{quote}
{\tt void upc\_memput(shared void *dst, const void *src, size\_t n);}
\end{quote}
will be used to transfer each outgoing message.

The meaning of {\tt mythread\_num\_recv\_values[T]}
 applies to the opposite communication direction.
Also,
the content of {\tt mythread\_recv\_value\_list[T]} will be needed by
{\tt MYTHREAD} to unpack the incoming message from thread {\tt T}.
One particular issue is that the {\tt upc\_memput} function requires a
pointer-to-shared available on the destination thread.
To this end, we need the following shared array with a block
size of {\tt THREAD}, where each array element is itself a pointer-to-shared:
\begin{quote}
{\tt shared[] double* shared [THREADS] shared\_recv\_buffers[THREADS*THREADS];}
\end{quote}
An important task in the preparation step is to let each thread go
through the following for-loop to allocate the individual buffers, in
UPC's globally shared address space, for its expected incoming messages:
\begin{quote}
\begin{verbatim}
for (int T=0; T<THREADS; T++)
  if (int length=mythread_num_recv_values[T]>0)
    shared_recv_buffers[MYTHREAD*THREADS+T]
     = (shared[] double*)upc_alloc(length*sizeof(double));
\end{verbatim}
\end{quote}
It should be noted that the standard
{\tt upc\_alloc} function should be called by
only one thread. The entire array that is allocated by {\tt upc\_alloc} has
affinity to the calling thread while being accessible by all the other
threads (see \cite{UPC_specification}). 
In the above for-loop, each thread (using its unique
{\tt MYTHREAD} value) only calls {\tt upc\_alloc} inside
its affinity block of {\tt shared\_recv\_buffers}.

\subsubsection{Communication procedure}

When the preparation step described above is done, we need to invoke
a communication procedure to precede each SpMV computation.
The communication procedure first lets each thread  (with {\tt
  MYTHREAD} as its unique ID) {pack} an
outgoing message for every thread {\tt T} that has 
{\tt mythread\_num\_send\_values[T]}$>0$, by extracting
the respective needed values from its owned blocks of the shared array
{\tt x} (cast to a pointer-to-local), 
using the local indices stored in {\tt mythread\_send\_value\_list[T]}.
Then, the one-sided communication function {\tt upc\_memput} is
called to send every ready-packed outgoing message to
its destination thread. Thereafter, the {\tt upc\_barrier}
command is posted to ensure that all the inter-thread communication is
done, which means that all the expected messages have arrived on
the respective destination
threads. Finally, each thread {unpacks} every incoming message by
copying its content to the respective positions in the thread-private 
array {\tt mythread\_x\_copy}. Each thread also copies its
 owned blocks from the shared array {\tt x} to the corresponding
 positions in the thread-private{\tt mythread\_x\_copy}.
The entire communication procedure can be seen in Listing~\ref{code-UPC-v3}.

\subsubsection{Implementation}

By incorporating the above preparation step and communication procedure,
we can create a new UPC implementation of SpMV in 
Listing~\ref{code-UPC-v3}. % The new version
% also adopts a private {\tt mythread\_x\_copy} array per thread,
%  as in the version of Listing~\ref{code-UPC-v2}.
% Its purpose is to acquire,
% before the SpMV, the needed $x$ values per thread so that accessing
% values through the pointer-to-shared {\tt x} is completely avoided in the
% actual SpMV computation.
% The difference between Listing~\ref{code-UPC-v3} 
% and Listing~\ref{code-UPC-v2} is that the new version adopts
% a more aggressive code transformation to condense and consolidate the
% between-thread messages.
Specifically, each pair of communicating threads
exchanges only one message containing the actually needed $x$
values. As a ``price'', the new version has to introduce
additional data structures in the preparation step,
and involve message packing and unpacking
in the communication procedure.

\begin{lstlisting}[caption=An improved UPC implementation
  of SpMV by message condensing and consolidation,
label=code-UPC-v3]
/* Allocation of the five shared arrays x, y, D, A, J as in the naive implementation */
// ...
/* Allocation of an additional private x array per thread */
double *mythread_x_copy = (double*) malloc(n*sizeof(double));

/* Preparation step: create and fill the thread-private arrays of int *mythread_num_send_values, int *mythread_num_recv_values, int **mythread_send_value_list, int **mythread_recv_value_list, double **mythread_send_buffers. Also, shared_recv_buffers is prepared. */
// ....

/* Communication procedure starts */
int T,k,mb,offset;
double *local_x_ptr = (double *)(x+MYTHREAD*BLOCKSIZE);
for (T=0; T<THREADS; T++)
  if (mythread_num_send_values[T]>0) /* pack outgoing messages */
    for (k=0; k<mythread_num_send_values[T]; k++)
      mythread_send_buffers[T][k] = local_x_ptr[mythread_send_value_list[T][k]];

for (T=0; T<THREADS; T++)
  if (mythread_num_send_values[T]>0) /* send out messages */
    upc_memput(shared_recv_buffers[T*THREADS+MYTHREAD], mythread_send_buffers[T],
               mythread_num_send_values[T]*sizeof(double));

upc_barrier;

int mythread_nblks = nblks/THREADS+(MYTHREAD<(nblks%THREADS)?1:0);
for (mb=0; mb<mythread_nblks; mb++) {/* copy own x-blocks */
  offset = (mb*THREADS+MYTHREAD)*BLOCKSIZE;
  memcpy(&mythread_x_copy[offset], (double *)(x+offset), min(BLOCKSIZE,n-offset)*sizeof(double));
}

for (T=0; T<THREADS; T++)
  if (mythread_num_recv_values[T]>0) {/* unpack incoming messages */
    double *local_buffer_ptr = (double *)shared_recv_buffers[MYTHREAD*THREADS+T];
    for (k=0; k<mythread_num_recv_values[T]; k++)
      mythread_x_copy[mythread_recv_value_list[T][k]] = local_buffer_ptr[k];
  }
/* Communication procedure ends */

/* SpMV: each thread only goes through its designated blocks */
for (mb=0; mb<mythread_nblks; mb++) {
  offset = (mb*THREADS+MYTHREAD)*BLOCKSIZE;
  /* casting shared pointers to local pointers */
  double *loc_y = (double*) (y+offset);
  double *loc_D = (double*) (D+offset);
  double *loc_A = (double*) (A+offset*r_nz);
  int *loc_J = (int*) (J+offset*r_nz);
  /* computation per block */
  for (k=0; k<min(BLOCKSIZE,n-offset); k++) {
    double tmp = 0.0;
    for (int j=0; j<r_nz; j++)
      tmp += loc_A[k*r_nz+j]*mythread_x_copy[loc_J[k*r_nz+j]];
    loc_y[k] = loc_D[k]*mythread_x_copy[offset+k] + tmp;
  }
}
\end{lstlisting}

%% file: Sec_PerfModels.tex
\section{Performance models}
\label{sec:perf_model}

We consider
the performance model of a parallel implementation as a formula
that can theoretically estimate the run time, based on some
information of the target work and some characteristic
parameters of the hardware platform intended.
Roughly, the time usage of a parallel program that implements a
computation comprises the time spent on the computational work and
the parallelization overhead. The latter is mostly spent on
various forms of communication between the executing processes or threads.

%For our specific example of computing a SpMV where each matrix row has
%a constant number of nonzeros, t
The three UPC implementations shown
in Section~\ref{sec:transforms} carry out identical computational
work. However, they differ greatly in
how the between-thread communication is realized, with respect to
both the frequency and volume of the beween-thread data transfers.
As will be demonstrated in Section~\ref{sec:experiments}, the time
usages of the three transformed UPC implementations are very
different. This motivates us to derive the corresponding
performance models, with a special focus on modeling the communication
cost in detail. Such theoretical performance models will help us to understand the
actual computing speed achieved, while also providing hints on further
performance tuning. %Moreover, .....

\subsection{Time spent on computation}
\label{sec:time_comp}

Due to a fixed number of nonzeros per matrix
row, the amount of floating-point operations per thread 
is linearly proportional to the number of $y(i)$ values that are
designated to each thread to compute. 
For all the UPC implementations in this paper, the shared array {\tt y} is
distributed in a block cyclic manner, with a programmer-prescribed block size of
{\tt BLOCKSIZE}. % Depending on the actual size of {\tt BLOCKSIZE},
% there can be some imbalance in the computational work division.
Recall that the array {\tt y} is of length $n$, thus
the number of {\tt y} blocks assigned per thread,
${B}^{\mathrm{comp}}_{\mathrm{thread}}$, is given by the following formula:
\begin{eqnarray}
{B}^{\mathrm{comp}}_{\mathrm{total}}&=&\Bigl\lceil \frac{n}{\mbox{{\tt BLOCKSIZE}}} \Big\rceil,
\nonumber\\
{B}^{\mathrm{comp}}_{\mathrm{thread}} &=& \Bigl\lfloor \frac{{B}^{\mathrm{comp}}_{\mathrm{total}}}{\mbox{{\tt THREADS}}} \Big\rfloor+
\left\{ \begin{array}{ll}
1&\mbox{ if } {\tt MYTHREAD}<({B}^{\mathrm{comp}}_{\mathrm{total}}\,
\mathrm{modulo}\mbox{ {\tt THREADS}}),\\
0&\mbox{ else.}
\end{array}
\right.
\label{eq:numblocks_thread}
\end{eqnarray}

% The performance bottleneck for scientific code usually lies
% inside the memory system of a modern computer, not in the
% floating-point units. 
Due to a low ratio between the number of
floating-point operations and the induced amount of data movement in the
memory hierarchy, the cost of computation for our SpMV example is
determined by the latter, as suggested by
 the well-known Roofline model~\cite{Williams:2009:RIV:1498765.1498785}.
Our strategy %to estimate the time spent on the computational work for each $y(i)$ value
is to derive the minimum amount of data movement needed between the main
memory and the last-level cache. 
More specifically, the following formula gives the minimum data
traffic (in bytes)
from/to the main memory for computing each $y(i)$ value:
\begin{eqnarray}
D_{\min}^{\mathrm{comp}}=r_{nz}\cdot(\mbox{{\tt
  sizeof(double)}}+\mbox{{\tt sizeof(int)}})
+3 \cdot\mbox{{\tt sizeof(double)}}.
\label{eq:min_memory_row}
\end{eqnarray}
Here, $r_{nz}$ denotes the fixed number of
off-diagonal nonzero values per matrix row, 
%where each nonzero value 
%{\tt A[i*r\_nz+j]}
%{\tt loc\_A[k*r\_nz+j]} occupies 
each occupying {\tt sizeof(double)} bytes
in memory, with {\tt sizeof(int)} bytes needed per column index.
% {\tt J[i*r\_nz+j]} 
%{\tt loc\_J[k*r\_nz+j]}. 
The last term in
 (\ref{eq:min_memory_row}) corresponds to the two memory loads
 for accessing
%{\tt D[i]} and {\tt x[i]} 
{\tt loc\_D[k]} and {\tt mythread\_x\_copy[offset+k]} (or {\tt x[offset+k]})
and the memory store associated with updating {\tt loc\_y[k]}. We refer to
Listings~\ref{code-UPC-v1}-\ref{code-UPC-v3} for the implementation details.

Formula (\ref{eq:min_memory_row}) has assumed perfect data reuse in
the last-level data cache. Our earlier experiences with the same SpMV
computation (implemented in sequential C or OpenMP), for the case
of a ``proper'' ordering of the matrix
rows (see e.g.~\cite{jpdc2015}), suggest 
% we have deliberately ignored the additional main-memory traffic that may
% occur with accessing the other $x$ values, apart from $x(i)$, needed
% for computing $y(i)$. The rationale is
that (\ref{eq:min_memory_row}) is a realistic estimate for 
 the last two UPC implementations
(Listings~\ref{code-UPC-v2} and \ref{code-UPC-v3}).
For these two implementations, the $x$ values
are fetched from the thread-private array {\tt mythread\_x\_copy}.
% for, data reuse of
% the $x$ values in the last-level cache is assumed to be sufficiently good.
% The minimum amount of data traffic from/to the main memory,
% quantified by (\ref{eq:min_memory_row}), will thus be quite close to the
% reality 
In the first transformed UPC implementation
(Listing~\ref{code-UPC-v1}), indirectly indexed accesses to the shared
array {\tt x} (of form {\tt x[loc\_J[k*r\_nz+j]]}) will incur additional memory
traffic on ``remote'' threads, caused by the inevitable between-thread data transfers.
We have chosen for this case to model 
the deviation from (\ref{eq:min_memory_row}) as a part of the
communication cost, to be discussed in Section~\ref{sec:5.2.3}.

% Based on (\ref{eq:numblocks_thread}), we can calculate the
% the number of $y$ values designated to a thread as
% $B^{\mathrm{comp}}_{\mathrm{thread}}\cdot\mbox{{\tt BLOCKSIZE}}$. We can thereby
% calculate the minimum amount of main-memory traffic associated with the thread's
% total SpMV computational work by multiplying with the value of
% $D_{\min}^{\mathrm{comp}}$ that is given by (\ref{eq:min_memory_row}).
Therefore, the minimum computational time needed per
thread, which is the same for all the UPC implementations of this paper, can be
estimated as
\begin{eqnarray}
T_{\mathrm{thread}}^{\mathrm{comp}} &=&\frac{B^{\mathrm{comp}}_{\mathrm{thread}}\cdot\mbox{{\tt
                      BLOCKSIZE}} \cdot
                      D_{\min}^{\mathrm{comp}}}{W_{\mathrm{thread}}^{\mathrm{private}}},
\label{eq:comp_time}
\end{eqnarray}
where $W_{\mathrm{thread}}^{\mathrm{private}}$ denotes the realistic
bandwidth (bytes per second) at which a thread can access its private memory space.
This can be found
by running a multi-threaded STREAM benchmark (see~\cite{STREAM}) on one
compute node of a target hardware platform, using the intended number of UPC
threads per node. The
$W_{\mathrm{thread}}^{\mathrm{private}}$ value equals the measured
multi-threaded STREAM bandwidth divided by the number of threads used.
Note that the bandwidth measured by a single-threaded STREAM
benchmark cannot be used directly as $W_{\mathrm{thread}}^{\mathrm{private}}$,
unless a single UPC thread per compute node is indeed intended.
This is because the multi-threaded STREAM bandwidth is
not linearly proportional to the %single-threaded STREAM bandwidth.
number of threads used, due to saturation of the memory bandwidth.

\subsection{Communication overhead}
\label{sec:time_comm}

\subsubsection{Definitions}
Before we dive into the details of modeling the various communication costs
that are associated with the three transformed UPC implementations, it
is important to establish the following definitions:
\begin{itemize}
\item If a thread accesses a memory location in
  the globally shared address space with affinity to another thread,
  a {\em non-private} memory operation is incurred.
\item A non-private memory operation, which is between two threads,
 can belong to one of two categories: {\em local inter-thread} and
 {\em remote
 inter-thread}. The first category refers to the case where the two
 involved threads reside on the same compute node, which has a
 physically shared NUMA (or UMA) memory encompassing all the threads
 running on the node. The second category refers to the case
 where the two threads reside on two different 
nodes, which need to use some interconnect for exchanging data.
\item A non-private memory operation, in each category, can happen in
  two modes:
  either individually or inside a sequence of memory
  operations accessing a contiguous segment of non-private
  memory. We term the first mode as {\em individual}, the second mode
  as {\em contiguous}.
\end{itemize}

\subsubsection{Cost of non-private memory operations}
\label{sec:non-priviate-mem}

The time needed by one non-private memory operation, in the
  contiguous mode, can be estimated as
\begin{eqnarray}
T^{\mathrm{local}}_{\mathrm{cntg}}=\frac{\mbox{{\tt
  sizeof}}\mbox{(one element)}}{W^{\mathrm{local}}_{\mathrm{thread}}}
\quad\mbox{ or }\quad
T^{\mathrm{remote}}_{\mathrm{cntg}}=\frac{\mbox{{\tt
  sizeof}}\mbox{(one element)}}{W^{\mathrm{remote}}_{\mathrm{node}}},
\label{eq:non-private}
\end{eqnarray}
where $W^{\mathrm{local}}_{\mathrm{thread}}$ denotes
the per-thread bandwidth for contiguous local inter-thread memory operations,
and we assume for simplicity
$W^{\mathrm{local}}_{\mathrm{thread}}=W^{\mathrm{private}}_{\mathrm{thread}}$,
with the latter being defined in Section~\ref{sec:time_comp}.
Correspondingly, $W^{\mathrm{remote}}_{\mathrm{node}}$ denotes
the interconnect bandwidth available to a node for contiguous remote (inter-node)
memory operations.
The reason for adopting a per-node bandwidth for inter-node memory
operations is because the inter-node network bandwidth can typically be
fully utilized by one thread, unlike the main-memory bandwidth.
The value of $W^{\mathrm{remote}}_{\mathrm{node}}$ can be measured
by a modified UPC STREAM benchmark or simply a standard MPI ping-pong
test, to be discussed in Section~\ref{sec:hw}.

%Modeling individual non-private memory operations requires some care. 
The cost of one individual remote
inter-thread memory operation, $T^{\mathrm{remote}}_{\mathrm{indv}}$,
is assumed to be dominated by a constant latency overhead, denoted
by $\tau$. Specifically, the latency $\tau$ is independent of the actual number
of bytes involved in one individual remote memory operation. By the
same reason $W^{\mathrm{remote}}_{\mathrm{node}}$ has no bearing on
$T^{\mathrm{remote}}_{\mathrm{indv}}$.
The actual value of $\tau$ can be measured by a special UPC
micro-benchmark, to be discussed in Section~\ref{sec:hw}.
The cost of one individual local inter-thread memory operation
can be estimated by the following formula:
\begin{eqnarray}
T^{\mathrm{local}}_{\mathrm{indv}} =\frac{\mbox{{\tt
  sizeof}}\mbox{(cache line)}}{W^{\mathrm{local}}_{\mathrm{thread}}}.
\label{eq:local_individual}
\end{eqnarray}
Here, we will again adopt
$W^{\mathrm{local}}_{\mathrm{thread}}=W^{\mathrm{private}}_{\mathrm{thread}}$. 
The reason for having the size of one cache line as the
numerator in (\ref{eq:local_individual})
 is that individual local inter-thread memory operations are
considered to be non-contiguously spread
in the private memory of the owner thread, thus
paying the price of an entire cache line per access.
(It has been implied that one data element occupies fewer bytes than
one cache line.)

\subsubsection{Communication time for the first transformed UPC implementation}
\label{sec:5.2.3}

For the UPC implementation in
Listing~\ref{code-UPC-v1}, an individual non-private memory
operation arises
% from the indirectly indexed accesses to the shared {\tt x} array.
when the owner thread of value {\tt x[loc\_J[k*r\_nz+j]]}
 is different from the accessing thread.
% a non-private memory access happens.
Each such non-private memory operation costs either
$T^{\mathrm{local}}_{\mathrm{indv}}$ as defined in (\ref{eq:local_individual})
or $T^{\mathrm{remote}}_{\mathrm{indv}}=\tau$. To quantify the
total communication time incurred per thread, we need the following
two counts, which can be obtained by letting each thread examine
its owned blocks of the shared array {\tt J}:
%by a thread (with a unique {\tt MYTHREAD} identity) 
% while each thread traversing its owned blocks of the {\tt J} array:
\begin{itemize}
\item
$C_{\mathrm{thread}}^{\mathrm{local,indv}}$:
Number of occurrences when {\tt
  \&x[loc\_J[k*r\_nz+j]]} has a different affinity than {\tt MYTHREAD}
and the
owner thread resides on the same compute node as {\tt MYTHREAD}.
\item
$C_{\mathrm{thread}}^{\mathrm{remote,indv}}$:
Number of occurrences when  {\tt
  \&x[loc\_J[k*r\_nz+j]]} has a different affinity than {\tt MYTHREAD}
and the
owner thread resides on a different compute node. % than that of {\tt MYTHREAD}.
\end{itemize}

Thus, the total commuication cost per thread during each SpMV is
\begin{eqnarray}
T_{\mathrm{thread}}^{\mathrm{comm,UPCv1}}
% = C_{\mathrm{thread}}^{\mathrm{local,indv}}\cdot
%   T^{\mathrm{local}}_{\mathrm{indv}}
% +C_{\mathrm{thread}}^{\mathrm{remote,indv}}\cdot
%   T^{\mathrm{remote}}_{\mathrm{indv}}
=C_{\mathrm{thread}}^{\mathrm{local,indv}}\cdot \frac{\mbox{{\tt
  sizeof}}\mbox{(cache line)}}{W^{\mathrm{private}}_{\mathrm{thread}}}
+C_{\mathrm{thread}}^{\mathrm{remote,indv}}\cdot\tau.
\label{eq:comm_upcv1}
\end{eqnarray}

\subsubsection{Communication time for the second transformed UPC
  implementation}
\label{sec:5.2.4}

For the UPC implementation in
Listing~\ref{code-UPC-v2}, before computing the SpMV, each
thread calls the {\tt upc\_memget} function to transport
 its needed blocks from the shared array {\tt x} to
the private array {\tt mythread\_x\_copy}.
To estimate the communication time spent per node, we will use the
following formula:
\begin{eqnarray}
T_{\mathrm{node}}^{\mathrm{comm,UPCv2}}
&=& \max_{\forall\, \mathrm{threads\;in\;node}}
B_{\mathrm{thread}}^{\mathrm{local}}\cdot\frac{2\cdot \mbox{{\tt
  BLOCKSIZE}}\cdot\mbox{{\tt sizeof(double)}
}}{W^{\mathrm{private}}_{\mathrm{thread}}} \nonumber \\
&+&
\sum_{\forall\, \mathrm{threads\;in\;node}}
B_{\mathrm{thread}}^{\mathrm{remote}}\cdot\left(\tau+\frac{\mbox{{\tt
  BLOCKSIZE}}\cdot\mbox{{\tt
  sizeof(double)}}}{W^{\mathrm{remote}}_{\mathrm{node}}}\right).
\label{eq:comm_upcv2_node}
\end{eqnarray}
Here, $B^{\mathrm{local}}_{\mathrm{thread}}$ denotes the number of
{\tt x} blocks residing on the same node as {\tt MYTHREAD} and
having at least one value needed by {\tt MYTHREAD},
 whereas $B^{\mathrm{remote}}_{\mathrm{thread}}$ denotes the number of
needed blocks residing on other nodes.
The reason for having a factor of 2 in the numerator of the first term
on the right-hand side of (\ref{eq:comm_upcv2_node}) is due to the
private/local memory loads and stores that both take place on the same node. 
Note that we have consistently assumed
$W^{\mathrm{local}}_{\mathrm{thread}}=W^{\mathrm{private}}_{\mathrm{thread}}$.
% The reason of not modeling the overhead associated with each thread is because
More importantly, we consider that all the threads on the same node
concurrently carry out their the intra-node part of communication,
whereas the inter-node operations of {\tt upc\_memput} are carried out one by one.
For communicating each inter-node block,
we have included $\tau$ as the ``start-up'' overhead
in addition to the $W^{\mathrm{remote}}_{\mathrm{node}}$-determined cost.

\subsubsection{Communication time for the third transformed UPC implementation}
\label{sec:5.2.5}

For the UPC implementation in
Listing~\ref{code-UPC-v3}, the overhead per thread for preparing the
private array {\tt mythread\_x\_copy} before the SpMV has four parts:
(1) packing all its outgoing messages; (2) calling {\tt
  upc\_memput} for each outgoing message; (3) copying its own blocks of
{\tt x} to the corresponding positions in {\tt mythread\_x\_copy}; (4)
unpacking the incoming messages.

Let us denote by $S_{\mathrm{thread}}^{\mathrm{local,out}}$ the
accumulated size of the outgoing messages from {\tt MYTHREAD} to threads
residing on the same node as {\tt MYTHREAD},
$S_{\mathrm{thread}}^{\mathrm{remote,out}}$ denotes the accumulated size of
the outgoing messages towards other nodes.
Similarly, $S_{\mathrm{thread}}^{\mathrm{local,in}}$
and $S_{\mathrm{thread}}^{\mathrm{local,in}}$ denote the incoming
counterparts.
Then, the per-thread overhead of packing the outgoing messages is
\begin{equation}
T^{\mathrm{pack}}_{\mathrm{thread}} =
\frac{(S_{\mathrm{thread}}^{\mathrm{local,out}}
  +S_{\mathrm{thread}}^{\mathrm{remote,out}})( 2\cdot\mbox{\tt
    sizeof(double)} +\mbox{\tt sizeof(int)})}
{W^{\mathrm{private}}_{\mathrm{thread}}}.
\label{eq:pack_cost}
\end{equation}
We remark that packing each value in an outgoing message requires loading at least
$\mbox{\tt sizeof(double)}+\mbox{\tt sizeof(int)}$ bytes from the private
memory, and storing $\mbox{\tt sizeof(double)}$ bytes into the message.

% The per-thread overhead of communicating the messages by {\tt upc\_memput} is
% \begin{equation}
% T^{\mathrm{memput}}_{\mathrm{thread}} =
% \frac{2\cdot S_{\mathrm{thread}}^{\mathrm{local,out}}\cdot \mbox{\tt
%     sizeof(double)}}{W^{\mathrm{private}}_{\mathrm{thread}}}
% +C_{\mathrm{thread}}^{\mathrm{remote,out}}\cdot\tau
% +\frac{S_{\mathrm{thread}}^{\mathrm{remote,out}}\cdot \mbox{\tt
%     sizeof(double)}}{W^{\mathrm{remote}}_{\mathrm{node}}},
% \label{eq:memput_cost}
% \end{equation}
% where $C_{\mathrm{thread}}^{\mathrm{remote,out}}$ denotes the number
% of outgoing inter-node messages from {\tt MYTHREAD}. 
Instead of modeling the per-thread overhead related to the {\tt
  upc\_memput} calls, we choose to model the per-node couterpart as
\begin{eqnarray}
T^{\mathrm{memput,UPCv3}}_{\mathrm{node}} &=&\max_{\forall\,\mathrm{threads\;in\;node}}
\frac{2\cdot S_{\mathrm{thread}}^{\mathrm{local,out}}\cdot \mbox{\tt
    sizeof(double)}}{W^{\mathrm{private}}_{\mathrm{thread}}} \nonumber\\
&+&\sum_{\forall\,\mathrm{threads\;in\;node}} \left(C_{\mathrm{thread}}^{\mathrm{remote,out}}\cdot\tau
+\frac{S_{\mathrm{thread}}^{\mathrm{remote,out}}\cdot \mbox{\tt
    sizeof(double)}}{W^{\mathrm{remote}}_{\mathrm{node}}}\right),
\label{eq:memput_cost_node}
\end{eqnarray}
where $C_{\mathrm{thread}}^{\mathrm{remote,out}}$ denotes the number
 of outgoing inter-node messages from {\tt MYTHREAD}.
Again, for each inter-node message, we have included $\tau$ as the ``start-up'' overhead
in addition to the $W^{\mathrm{remote}}_{\mathrm{node}}$-determined cost.

The per-thread overhead of copying the private blocks of {\tt x} into {\tt
  mythread\_x\_copy} is
 \begin{equation}
T^{\mathrm{copy}}_{\mathrm{thread}} =
\frac{2\cdot B^{\mathrm{comp}}_{\mathrm{thread}}\cdot \mbox{\tt
    BLOCKSIZE}\cdot \mbox{\tt
    sizeof(double)}}{W^{\mathrm{private}}_{\mathrm{thread}}},
\label{eq:copy_cost}
\end{equation}
where we recall that $B^{\mathrm{comp}}_{\mathrm{thread}}$ is defined
in (\ref{eq:numblocks_thread}).

Finally, the per-thread overhead of unpacking the incoming messages is
\begin{equation}
T^{\mathrm{unpack}}_{\mathrm{thread}} =
\frac{(S_{\mathrm{thread}}^{\mathrm{local,in}}
  +S_{\mathrm{thread}}^{\mathrm{remote,in}})(\mbox{\tt
    sizeof(double)} +\mbox{\tt sizeof(int)} + \mbox{{\tt sizeof}(cache
    line)})}
{W^{\mathrm{private}}_{\mathrm{thread}}}.
\label{eq:unpack_cost}
\end{equation}
Note that $\mbox{\tt
    sizeof(double)} +\mbox{\tt sizeof(int)}$ corresponds to
  contiguously reading each value from an incoming message, whereas
$\mbox{{\tt sizeof}(cache line)}$ corresponds to the cost of writing
the value to a non-contiguous location in the array {\tt mythread\_x\_copy}.

% The total overhead of communication is thus
% \begin{eqnarray}
% T_{\mathrm{thread}}^{\mathrm{comm,UPCv3}}=
% T^{\mathrm{pack}}_{\mathrm{thread}} +
% T^{\mathrm{memput}}_{\mathrm{thread}} +
% T^{\mathrm{copy}}_{\mathrm{thread}}  +
% T^{\mathrm{unpack}}_{\mathrm{thread}}.
% \label{eq:comm_upcv3}
% \end{eqnarray}

\subsection{Total time usage}

Due to the possible imbalance of both computational work and
communication overhead among the threads,
the total time usage of any of the UPC implementations will be
determined by the slowest thread or node. For the first transformed
UPC implementation, shown in Listing~\ref{code-UPC-v1},  the total
time is determined by the slowest thread:
\begin{eqnarray}
T_{\mathrm{total}}^{\mathrm{UPCv1}}=\max_{\forall \,\mathrm{threads}}
  \left(T_{\mathrm{thread}}^{\mathrm{comp}}+T_{\mathrm{thread}}^{\mathrm{comm,UPCv1}}\right).
\end{eqnarray}

For the second transformed
UPC implementation, shown in Listing~\ref{code-UPC-v2}, the total time is
determined by the slowest node:
\begin{eqnarray}
T_{\mathrm{total}}^{\mathrm{UPCv2}}=\max_{\forall \,\mathrm{nodes}}
  \left(\left(\max_{\forall \,\mathrm{threads\;in\;node}}
  T_{\mathrm{thread}}^{\mathrm{comp}}\right)+T_{\mathrm{node}}^{\mathrm{comm,UPCv2}}\right).
\end{eqnarray}

For the third transformed UPC implementation, shown in
Listing~\ref{code-UPC-v3}, due to the needed explicit barrier
after the {\tt upc\_memput} calls, the total time usage is modeled as
\begin{eqnarray}
T_{\mathrm{total}}^{\mathrm{UPCv3}}=\max_{\forall\,\mathrm{nodes}}
\left(\left(\max_{\forall \, \mathrm{threads\;in\;node}}T_{\mathrm{thread}}^{\mathrm{pack}}\right)
+T_{\mathrm{node}}^{\mathrm{memput,UPCv3}}\right)
+\max_{\forall \, \mathrm{threads}}
  \left(T_{\mathrm{thread}}^{\mathrm{copy}}+T_{\mathrm{thread}}^{\mathrm{unpack}}
+T_{\mathrm{thread}}^{\mathrm{comp}}\right).
\end{eqnarray}

\subsection{Remarks}

%Should add remarks that are important for a reader to easily grasp the rationale and main
%concepts of the performance models....
 
It is important to separate two types of information needed by the
above performance models. The hardware-specfic information includes
$W_{\mathrm{thread}}^{\mathrm{private}}$,
$W_{\mathrm{node}}^{\mathrm{remote}}$, $\tau$ and the cache line
size of the last-level cache. The first
parameter denotes the per-thread rate of contiguously accessing
private memory locations. The second parameter is the per-node
counterpart for contiguously accessing remote off-node memory locations.
Note that we do not distinguish between $W_{\mathrm{thread}}^{\mathrm{private}}$
and intra-socket or inter-socket local memory bandwidths, due to
very small differences between them.
The $\tau$ parameter describes the latency for
an individual remote memory access. All the hardware parameters are
easily measurable by simple benchmarks, see Section~\ref{sec:hw}, or
known from hardware specification.

The computation-specific information includes
$C^{\mathrm{local,indv}}_{\mathrm{thread}}$,
$C^{\mathrm{remote,indv}}_{\mathrm{thread}}$ (Section~\ref{sec:5.2.3}), 
$B^{\mathrm{local}}_{\mathrm{thread}}$,
$B^{\mathrm{remote}}_{\mathrm{thread}}$ (Section~\ref{sec:5.2.4}), and
$S^{\mathrm{local,out}}_{\mathrm{thread}}$, $S^{\mathrm{local,in}}_{\mathrm{thread}}$,
$C^{\mathrm{remote,out}}_{\mathrm{thread}}$, $S^{\mathrm{remote,out}}_{\mathrm{thread}}$,
$S^{\mathrm{remote,in}}_{\mathrm{thread}}$ (Section~\ref{sec:5.2.5}).
These numbers depend on the specific spread of the nonzero values in
the sparse matrix. They can be obtained by letting each thread go
through its owned blocks of the shared array {\tt J} and do an
appropriate counting. Another important input is the programmer-chosen
value of {\tt BLOCKSIZE}, which controls how all the shared arrays are
distributed among the threads, thus determining all the above
computation-specific parameters.

%competition of resources

% ... should perhaps also add a nomenclature table listing the meaning of each
% symbol used in the different performance models.

%% file: Sec_Experiments.tex
\section{Experiments}
\label{sec:experiments}

To study the impact of various code transformations described in
Section~\ref{sec:transforms} and to validate the corresponding performance
models proposed in Section~\ref{sec:perf_model}, we will use a
real-world case of SpMV in this section.

\subsection{A 3D diffusion equation solver based on SpMV}
\label{sec:diff3D}

One particular application of SpMV can be found in numerically solving a 3D diffusion
equation that is posed on an irregular domain. Typically, an unstructured computational
mesh must be used to match the irregular domain.
All numerical strategies will involve a time integration process.
During time step $\ell$, the simplest numerical strategy takes the form of an SpMV:
\begin{eqnarray*}
v^{\ell}=Mv^{\ell-1},
\end{eqnarray*}
where vectors $v^{\ell}$ and $v^{\ell-1}$ denote the numerical
solutions on two consecutive time levels, each containing
approximate values on some mesh entities (e.g.~the centers of all tetrahedrons).
The $M$ matrix arises from a numerical discretization of the original
diffusion equation. Matrix $M$ is normally time-independent and thus
computed once and for all, prior to the time integration process. 
The unstructured computational mesh will lead to an irregular spread of the nonzeros.
Particularly, if a second-order finite volume discretization is applied to a
tetrahedral mesh, the number of off-diagonal nonzero values per row of
$M$ is up to 16, see e.g.~\cite{micro2015}.

Three test problems of increasing resolution will be used in this
section. They all arise from modeling the left cardiac ventricle of a
healthy male human.
(The 3D diffusion solver can be an integral part of a heart simulator.)
The three corresponding tetrahedral meshes are generated by
the open-source {\em TetGen} software~\cite{tetgen}, with the actual size
of the meshes being listed in
Table~\ref{tab:test_problem_size}. Note that we have $r_{nz}=16$ for all the
three test problems. The tetrahedrons have been re-ordered in each mesh for
achieving good cache behavior associated with a straightforward sequential
computation. It is important to notice that all the three meshes are
fixed for the following UPC computations, independent of the number of
UPC threads used and the value of {\tt BLOCKSIZE} chosen.

\begin{table}[ht]
\centering
\caption{Size of the three test problems.}
\label{tab:test_problem_size}
\begin{tabular}{l|c|c|c}
\hline
&Test problem 1&Test problem 2&Test problem 3\\
\hline
Number of tetrahedrons: $n$& 6,810,586&13,009,527&25,587,400\\
\hline
\end{tabular}
\end{table}

For any computer program implementing the 3D diffusion solver, two
arrays are sufficient
 for containing the two consecutive numerical solutions $v^{\ell}$ and $v^{\ell-1}$.
For the UPC implementations discussed in
Section~\ref{sec:transforms}, the shared array {\tt y} corresponds to $v^{\ell}$ and
{\tt x} to $v^{\ell-1}$ during each time step. The pointers-to-shared {\tt y} and {\tt x}
need to be swapped before the next time step, fenced between a pair of {\tt
  upc\_barrier} calls.

\input{Subsec_HW_platform.tex}

\input{Subsec_Measurement.tex}

\subsection{Validating the performance models}
\label{sec:model_validation}

We have seen in Section~\ref{sec:measurements} 
that the three transformed implementations
have quite different performance behaviors. To shed some light on
the causes of the performance differences, we will now
use the hardware characteristic parameters 
obtained in Section~\ref{sec:hw} together with the performance models
proposed in Section~\ref{sec:perf_model}. Specifically,
Table~\ref{tab:comparison} compares the actual time measurements
against the predicted time usages for Test problem 1 on the Abel cluster.
It can be seen that the predictions made by the performance models
of Section~\ref{sec:perf_model} follow the same trends of the actual time
measurements, except for the case of
UPCv1 using 128 threads. It is worth noticing that the single-node performance (16
threads) of UPCv2 is correctly predicted to be slower than that of
UPCv1, whereas the reverse of the performance relationship when using
multiple nodes is also
confirmed by the predictions.
For small thread counts (16--64), the prediction
accuracy is quite good. For larger threads counts, the predictions become
less accurate.

 For UPCv1, there are four cases where the actual run times are faster
 than the predictions. These are attributed to the fact that the
 adopted $\tau$ value of 3.4$\mu$s can be
 a little ``pessimistic''. Recall from Section~\ref{sec:hw} that the
 particular $\tau$ value was measured by the micro-benchmark when it used
 8 threads on one node to simultaneously communicate with 8 other threads
 on another node. In reality, the effective $\tau$ value can be smaller
 than 3.4$\mu$s, if the average number of remotely-communicating
 threads per node over time is fewer than 8.
For UPCv3, there are two cases where the actual run times are slightly faster
 than the predictions. This is due to imbalance between the
 threads with respect to the per-thread amount of computation
 and message packing/unpacking. When most of the threads have finished their
 tasks, the remaining threads will each have access to an effective
  $W_{\mathrm{thread}}^{\mathrm{private}}$ value that is larger than
  $\frac{1}{16}$ of $W_{\mathrm{node}}^{\mathrm{private}}$. This can
  result in the time prediction of UPCv3 being a little ``pessimistic''.

\begin{table}[ht]
\centering
\caption{Comparison between actual and predicted time usages (in
  seconds) of the three transformed UPC implementations for Test
  problem 1 ($n=6810586$). The hardware characteristic parameters used for the Abel
  cluster (16 UPC threads per node) are
  $W_{\mathrm{thread}}^{\mathrm{private}}=\frac{75\mathrm{GB/s}}{16}$,
$W^{\mathrm{remote}}_{\mathrm{node}}=6\mathrm{GB/s}$, $\tau=3.4\mu\mathrm{s}$.
}
\label{tab:comparison}
\begin{tabular}{c|c||c|c||c|c||c|c}
\hline
%%&\multicolumn{2}[||c}{UPC V1}&\multicolumn{2}[||c}{UPC
%%                               V2}&\multicolumn{2}[||c}{UPC V3}\\
%%\hline
{\small\tt THREADS}&{\small\tt BLOCKSIZE}&$T^{\mathrm{UPCv1}}_{\mathrm{total,actual}}$
&$T^{\mathrm{UPCv1}}_{\mathrm{total,predicted}}$
&$T^{\mathrm{UPCv2}}_{\mathrm{total,actual}}$
&$T^{\mathrm{UPCv2}}_{\mathrm{total,predicted}}$
&$T^{\mathrm{UPCv3}}_{\mathrm{total,actual}}$
&$T^{\mathrm{UPCv3}}_{\mathrm{total,predicted}}$\\
\hline
16&65536&28.80&26.40&39.37&37.21&25.01&22.95\\
32&65536&522.15&410.86&36.70&34.30&15.07&14.07\\
64&65536&443.98&607.08&23.68&20.19&8.22&7.83\\
128&53200&1882.01&677.99&18.89&12.43&4.65&4.07\\
256&26600&551.20&679.83&13.61&9.59&2.91&3.06\\
512&13300&311.54&388.42&9.98&7.83&2.68&2.96\\
1024&6650&183.73&200.96&9.57&8.15&5.56&3.55\\
\hline
\end{tabular}
\end{table}

\begin{figure}[htb]
\centerline{{\includegraphics[width=15cm, trim=0.78in 3.165in 0.93in
    3.18in, clip]{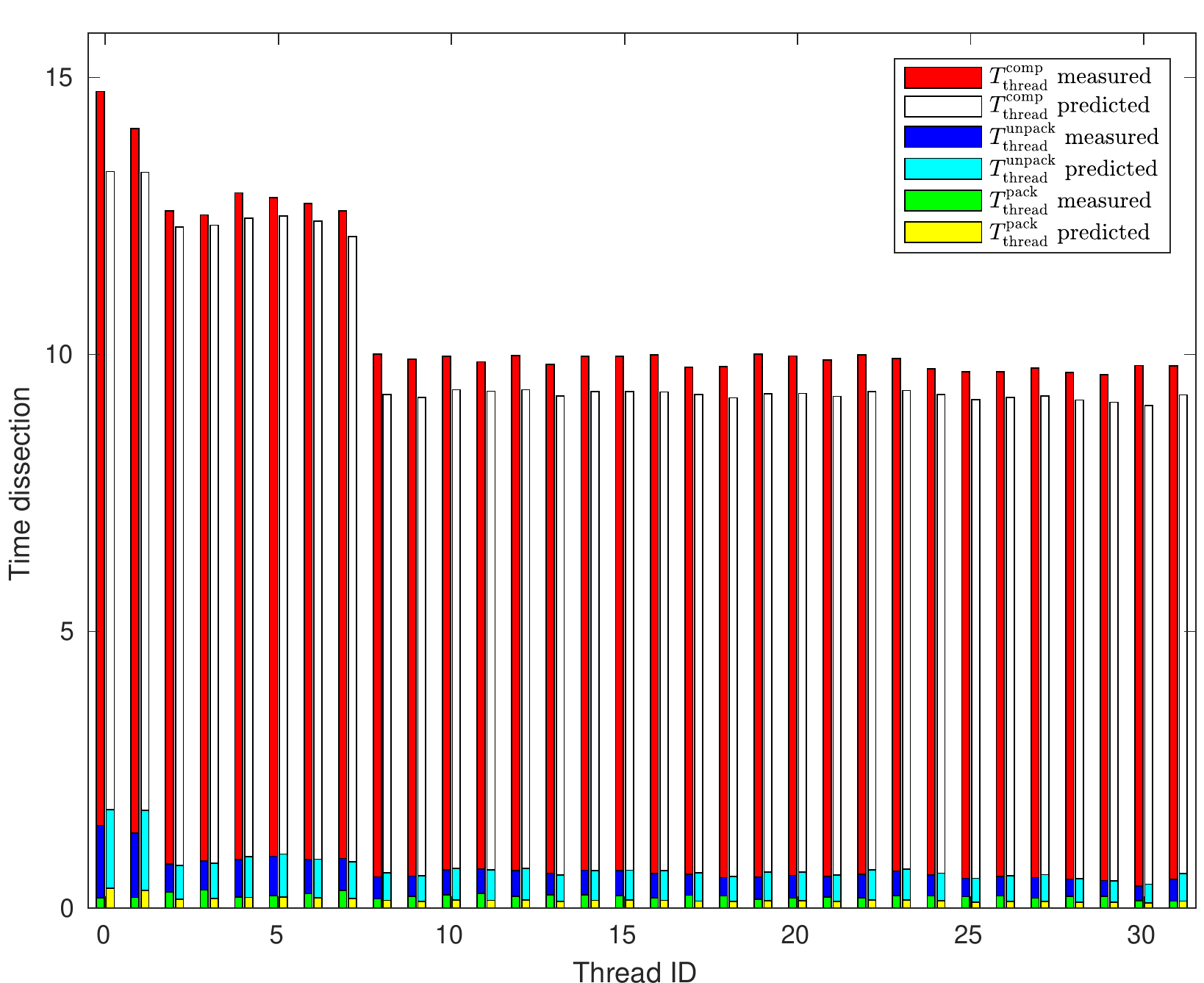}}}
\caption{Comparison between per-thread predictions and measurements of
$T^{\mathrm{comp}}_{\mathrm{thread}}$,
$T^{\mathrm{unpack}}_{\mathrm{thread}}$ and
$T^{\mathrm{pack}}_{\mathrm{thread}}$ for UPCv3;
Test problem 1 ($n=6810586$), 32 threads spread
  over two nodes, with {\tt BLOCKSIZE}=65536.}
\label{fig:time_dissect_V3}
\end{figure}

To examine some of the prediction details of UPCv3, we show in
Figure~\ref{fig:time_dissect_V3} the per-thread measurements and predictions of
$T^{\mathrm{comp}}_{\mathrm{thread}}$,
$T^{\mathrm{unpack}}_{\mathrm{thread}}$,
$T^{\mathrm{pack}}_{\mathrm{thread}}$ (see Section~\ref{sec:5.2.5}), for the particular case of
using 32 threads on two nodes. It can be seen that the
predictions of the three time components closely match the actual time
usages.

\begin{figure}[p]
\centerline{{\includegraphics[width=13.5cm, trim=0.78in 3.165in 0.93in
    3.18in, clip]{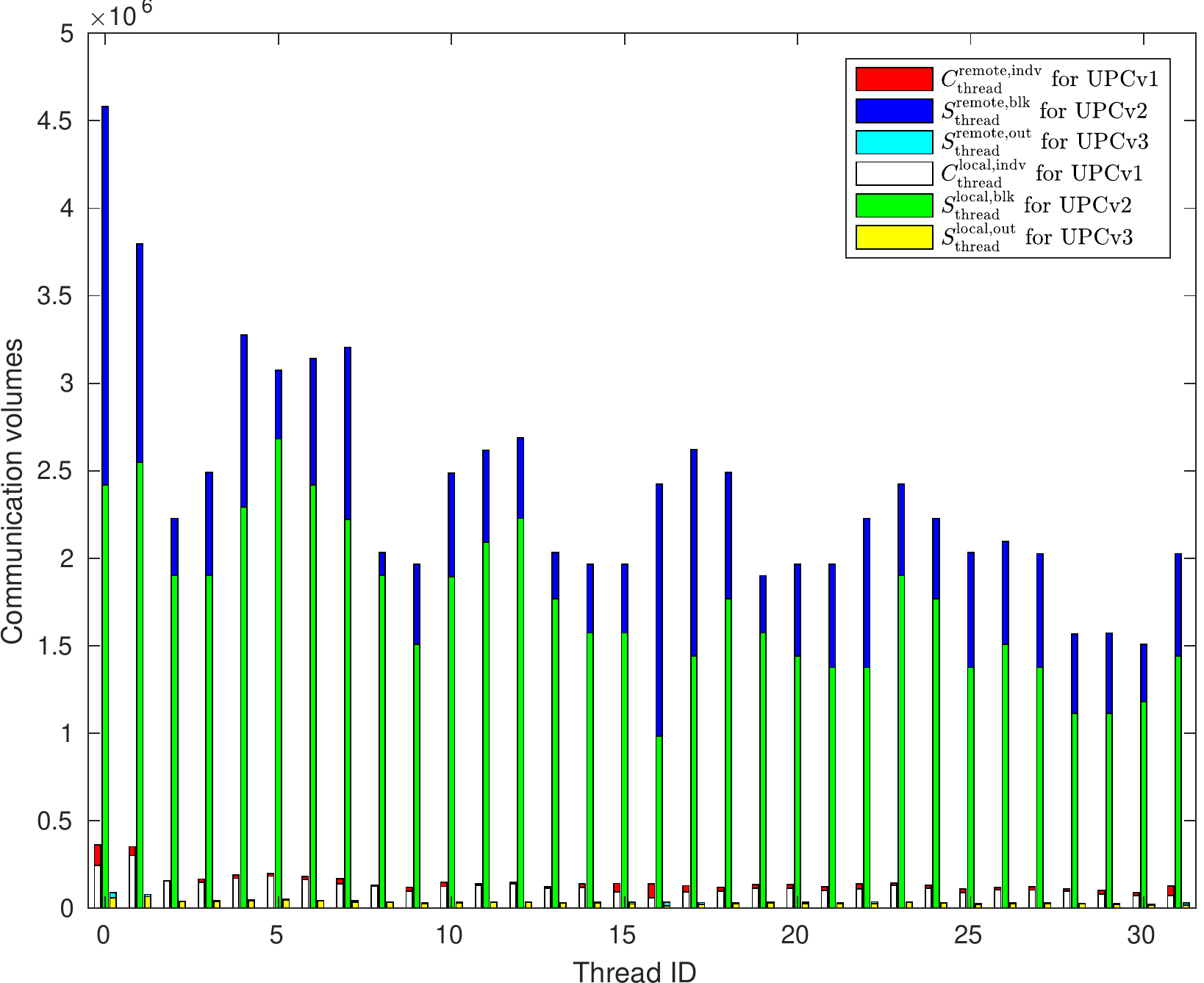}}}
\centerline{{\includegraphics[width=13.5cm, trim=0.78in 3.165in 0.93in
    3.18in, clip]{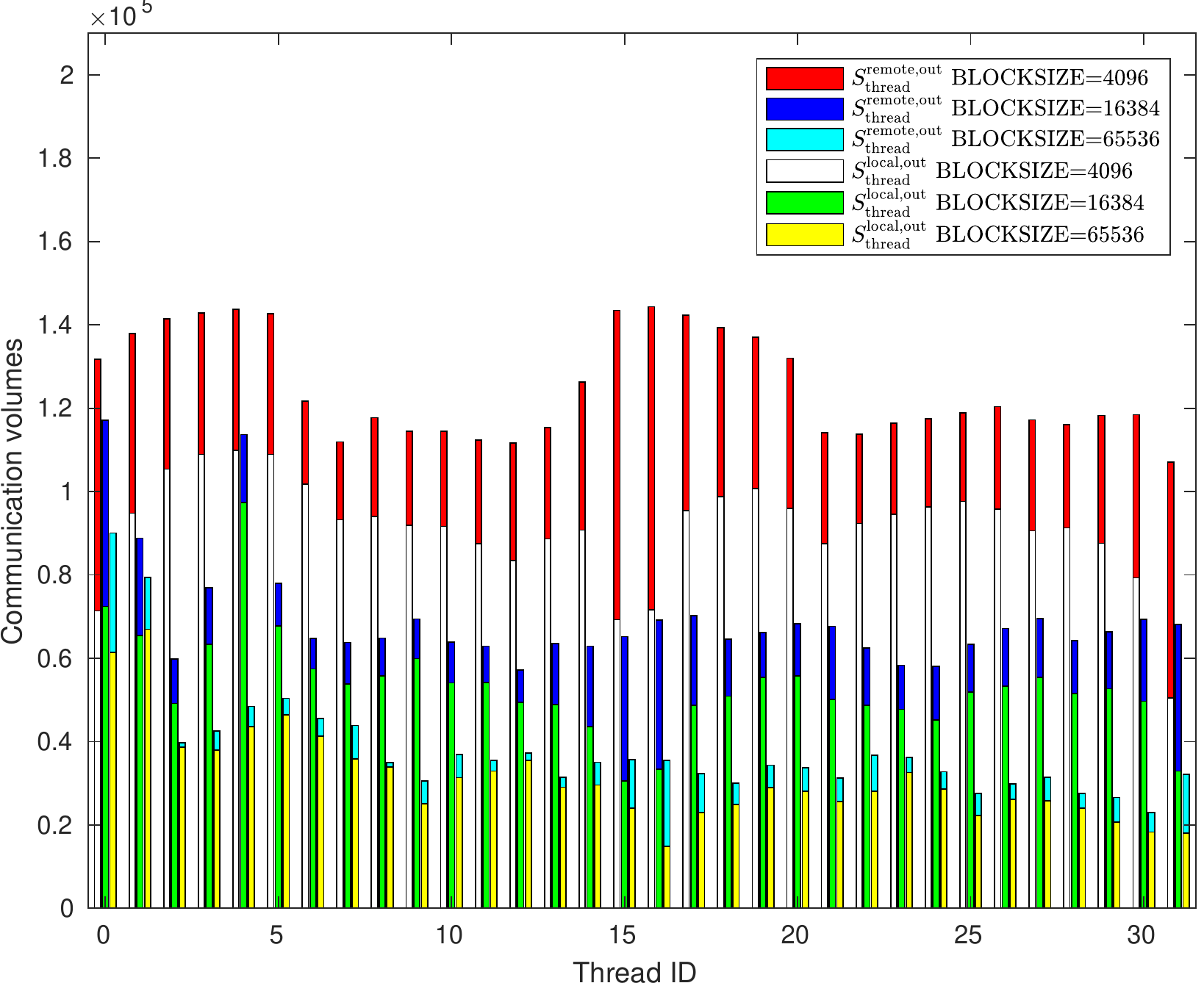}}}
\caption{Test problem 1 ($n=6810586$), 32 threads spread
  over two nodes. Top: Per-thread communication volumes required by the three
  transformed UPC implementations with {\tt BLOCKSIZE}=65536. Bottom:
  Per-thread communication volumes associated with UPCv3 for different values of
{\tt BLOCKSIZE}.}
\label{fig:comm_volumes}
\end{figure}

As mentioned in Section~\ref{sec:transforms}, the three UPC
implementations differ in how the inter-thread communications are
handled. To clearly show the difference, we have
plotted in the top of Figure~\ref{fig:comm_volumes} the per-thread
distribution of communication volumes for the specific case of using
32 threads with {\tt BLOCKSIZE} set to 65536. We observe
that UPCv3 has the lowest communication volume,
whereas UPCv2 has the highest. Although UPCv1 induces lower
communication volumes than UPCv2, all communications of the former are
individual and thus more costly. It is also observed that the
communication volumes can vary considerably from thread to thread.
The specific variation depends on the spread of the
nonzeros, as well as the number of threads used and the value of {\tt
  BLOCKSIZE} chosen. The dependency on the last factor is exemplified
in the bottom plot of Figure~\ref{fig:comm_volumes}. This shows that
tuning {\tt BLOCKSIZE} by the programmer is a viable approach to
performance optimization. The performance models are essential in this context.

% \begin{itemize}
% \item At least one table showing the accuracy of the three performance
%   models against the actual time measurements of UPC V1, V2, V3;
% \item At least one graph comparing the total communication volumes per
%   thread of UPC V1, V2, V3;
% (Use e.g. the case of using 32 threads and the D67 problem size. 
% The purpose is to show the inherent advantages/disadvantages of the
% three UPC implementations.)
% \item One graph showing the ``impact'' of tuning {\tt
%     BLOCKSIZE} on the frequency and volume of both
% local and remote inter-thread communications. 
% % Show different examples of data and work division, arisen from different
% % values of {\tt BLOCKSIZE}, and different total numbers of threads... 
% {\sl This is an important point of this paper, showing the very limited
% flexibility of UPC with respect to controlling the communication overhead.}
% (Choose 3 or 4 typical values of {\tt
%     BLOCKSIZE}, fixed on using e.g. the D67 problem
%   size. The purpose is to show that the resulting performances differ
%   quite much.)
% \item At least one graph dissecting the time usage of UPC V3 (actual
%   vs.~predicted time usages of computation, packing, upc\_memput,
%   unpacking).
% \end{itemize}

%% file: Subsec_HW_platform.tex
\subsection{Hardware and software platforms}
\label{sec:hw}

The Abel computer cluster~\cite{abel_cluster} was used to run all the
UPC codes and measure their time usage. Each compute node on Abel
is equipped with two Intel Xeon E5-2670 2.6~GHz 8-core CPUs and
 64~GB of RAM. The interconnect between the nodes is FDR InfiniBand 
(56~Gbits/s). With a multi-threaded STREAM~\cite{STREAM}
 micro-benchmark in C, we measured the aggregate memory
 bandwidth per node as 75~GB/s using 16 threads. This gave
 $W^{\mathrm{private}}_{\mathrm{thread}}=\frac{75}{16}$~GB/s. The
 inter-node communication bandwidth,
 $W^{\mathrm{remote}}_{\mathrm{node}}$ (defined in
 Section~\ref{sec:non-priviate-mem}), was measured by a standard MPI
 pingpong micro-benchmark to be about 6~GB/s.

The Berkeley UPC~\cite{Berkeley_UPC} version 2.24.2 was used for
compiling and running all our UPC implementations of SpMV. The compilation procedure
involved first a behind-the-scene translation from UPC to C done remotely
 at Berkeley via HTTP, with the translated C code being then compiled
locally on Abel using Intel's {\tt icc} compiler version
15.0.1. The compilation options were {\tt -O3 -wd177 -wd279 -wd1572 -std=gnu99}.

% When running our experiments we always used thread-binding. Thus, threads do not migrate from one core to another and the data related to those thread is placed in memory in a fixed way.

In order to measure the cost of an individual remote memory transfer, 
$\tau$ (defined in Section~\ref{sec:non-priviate-mem}), we developed a
micro-benchmark shown in Listing~\ref{code-tau}.
Specifically, {\tt v} is a shared UPC array created by {\tt upc\_all\_alloc}. Each
thread then randomly reads entries of {\tt v} that have affinity with
``remote threads'', through a thread-private index array {\tt
  mythread\_indices}.
The total time usage, subtracting the time needed to contiguously
traverse {\tt mythread\_indices}, can be used to quantify $\tau$.
When using two nodes each running 8 UPC threads, we measured the value
of $\tau$ as 3.4$\mu$s. Varying the number of concurrent threads does not
change the measured value of $\tau$ very much.

\begin{lstlisting}[caption=A UPC micro-benchmark for measuring
  the latency of individual remote memory transfers,label=code-tau]
int nblks = n/BLOCKSIZE + (n%BLOCKSIZE)?1:0;
int mythread_nblks = nblks/THREADS + (MYTHREAD<(nblks%THREADS)?1:0);
shared [BLOCKSIZE] double *v = upc_all_alloc (nblks, BLOCKSIZE*sizeof(double));
double tmp;
int *mythread_indices =(int*)malloc(mythread_nblks*BLOCKSIZE*sizeof(int));
/* let array 'mythread_indices' contain random global indices with affinity to ``remote threads'' */
randomize (mythread_indices, mythread_nblks*BLOCKSIZE);
/* start timing .... */
for (int mb=0,i=0; mb<mythread_nblks; mb++)
   for (int k=0; k<BLOCKSIZE; k++,i++)
      tmp = v[mythread_indices[i]];
/* stop timing .... */
\end{lstlisting}

% \begin{itemize}
% \item
% Description of the Abel cluster %and perhaps also the in-house KNL machine.
% \item
% UPC compiler and compilation options used
% \item
% Value of measured $W_{\mathrm{thread}}^{\mathrm{private}}$  value (by
% using plain C version of STREAM?)
% \item
% Value of measured $W^{\mathrm{remote}}_{\mathrm{node}}$
% (simply using an existing UPC benchmark from Ohio State?)
% \item
% UPC benchmark for measuring $\tau$ (cost of one individual remote memory access)
% \end{itemize}

%% file: Subsec_Measurement.tex
\subsection{Time measurements}
\label{sec:measurements}

Table~\ref{tab:one_node} compares the performance of the naive UPC
implementation (Listing~\ref{code-UPC-naive}) against that of
the first transformed UPC implementation (Listing~\ref{code-UPC-v1}).
%The latter will later be denoted by UPCv1.
Here, we only used one compute node on Abel while varying the number of UPC
threads. Each experiment was repeated several times, and the
best time measurement is shown in Table~\ref{tab:one_node}.
 Thread binding was always used, as for all the subsequent experiments.
Test problem 1 (with 6810586 tetrahedrons) was chosen with the value
of {\tt BLOCKSIZE} being fixed at 65536.
We can clearly see from Table~\ref{tab:one_node} that the naive UPC
implementation is very ineffective due to using {\tt upc\_forall} and
accessing {\tt y}, {\tt D}, {\tt A} and {\tt J} through pointers-to-shared.

\begin{table}[htb]
\caption{Time usage (in seconds) of 1000 iterations SpMV for Test
  problem 1; naive UPC
  implementation (Listing~\ref{code-UPC-naive}) vs.~the first transformed UPC
  implementation (Listing~\ref{code-UPC-v1}).}
\label{tab:one_node}
\centering
\begin{tabular}{l||c|c|c|c|c}
\hline
&1 thread&2 threads&4 threads&8 threads&16 threads\\
\hline
Naive UPC&895.44&548.57&301.17&173.08&106.10\\
UPCv1&270.40&159.51&86.37&51.10&28.80\\
\hline
\end{tabular}
\end{table}

Table~\ref{tab:multi_nodes} summarizes the time measurements for all
the three transformed UPC implementations (denoted by UPCv1,2,3) and all the three test
problems. It can be seen that UPCv3 has the best performance as expected, followed
by UPCv2 with UPCv1 being the slowest. The only exception is that
UPCv1 is faster than UPCv2 when running 16 UPC threads on one Abel
node. This is because there is no ``penalty'' of individual remote
memory transfers for UPCv1 in
such a scenario, whereas UPCv2 has to transfer all the needed blocks in entirety.

\begin{table}[htb]
\caption{Time usage (in seconds) of 1000 iterations SpMV for Test
  problems 1-3.}
\label{tab:multi_nodes}
\centering
\begin{tabular}{l||c|c|c|c|c|c|c}
\hline
&1 node&2 nodes&4 nodes&8 nodes&16 nodes&32 nodes&64 nodes\\
&16 threads&32 threads&64 threads&128 threads&256 threads&512
                                                           threads&1024
  threads\\
\hline
\multicolumn{7}{c}{Test problem 1: 6,810,586 tetrahedrons}\\
\hline
UPCv1&28.80&522.15&443.98&1882.01&551.20&311.54&183.73\\
UPCv2&39.37&36.70&23.68&18.89&13.61&9.98&9.57\\
UPCv3&25.01&15.07&8.22&4.65&2.91&2.68&5.56\\
\hline
\multicolumn{7}{c}{Test problem 2: 13,009,527 tetrahedrons}\\
\hline
UPCv1&59.14&2525.05&3532.33&3657.95&3078.35&2613.85&1588.67\\
UPCv2&73.79&69.60&55.33&36.39&24.16&25.06&21.29\\
UPCv3&46.88&24.97&15.43&10.91&6.25&5.15&7.54\\
\hline
\multicolumn{7}{c}{Test problem 3: 25,587,400 tetrahedrons}\\
\hline
UPCv1&115.25&2990.92&1758.94&986.85&1302.52&4653.10&2692.69\\
UPCv2&154.72&178.14&122.38&81.77&52.99&41.16&44.80\\
UPCv3&93.30&48.74&26.13&15.37&11.12&7.41&10.16\\
\hline
\end{tabular}
\end{table}

%% file: Sec_Related_work.tex
\section{Related work}
\label{sec:related_work}

Many performance studies about UPC programming,
e.g.~\cite{Ghazawi2002,Zhang-IPDPS2005,Barton2007,Shan2010}, selected kernels
from the NAS parallel benchmark (NPB) suite~\cite{NAS_techreport}. These
studies however did
not involve irregular fine-grained communication that arises from indirectly
indexing the elements of shared arrays. Other published non-NPB benchmarks implemented
in UPC, such as reported in~\cite{Barton2007}, had the same
limitation. Various UPC implementations of SpMV were studied
in~\cite{Dominguez2013}, but the authors chose to combine a row-wise
block distribution of the sparse matrix with duplicating the entire source vector $x$ on
each thread. Such a UPC implementation of SpMV completely avoided
the impact of irregular fine-grained communication, which is the main focus of
our present paper. The authors of~\cite{Li2015} distributed the source vector $x$
among the UPC threads, but their UPC implementation of SpMV explicitly
avoided off-node irregular fine-grained communication, for which the needed values
of $x$ were transported in batches from the off-node owner threads
using e.g.~{\tt upc\_memget}.

An extended suite of UPC STREAM micro-benchmarks, including various
scenarios of using pointers-to-shared, was proposed
by the authors of~\cite{Zhang-IPDPS2005}. They also reported measurements
that clearly reveal the additional overhead due to UPC language features.
For our purpose of performance modeling, we only found the so-called
``random shared read'' micro-benchmark defined
in~\cite{Zhang-IPDPS2005} to be useful for quantifying $\tau$.
This prompted us to write our own micro-benchmark (see
Section~\ref{sec:hw}) because we have no access to the source code
used for~\cite{Zhang-IPDPS2005}.

One approach to alleviating the various types of overhead associated
with using pointers-to-shared is by specialized UPC compilers, 
see e.g.~\cite{Cantonnet-IPDPS2005,Chen2005,Chen_PhD2007,Alvanos_PhD2013}.
However, indirectly accessing array elements through
pointers-to-shared, which will induce irregular fined-grained
communication, can not be handled by any of the specialized UPC
compilers. Manual code transformations are thus necessary, such as those proposed in
Sections~\ref{sec:4.2}-\ref{sec:4.3} of our present paper.

It is quite common that performance studies (such
as~\cite{Barton2007}) and performance models (such as~\cite{Zhang-IPDPS2006})
of UPC programs are built upon ``single-value statistics'' that is accumulated
or averaged over all threads.
We see this as an unignorable source of inaccuracy, because considerable variations in
the communication volume and pattern may exist between threads, as
exemplified by Figures~\ref{fig:time_dissect_V3} and \ref{fig:comm_volumes}.
Ignoring such thread-wise imbalances in communication
 will lead to inaccurate performance predictions, exactly in the same
 way as ignoring thread-wise imbalances in computation.
At the same time, the performance models proposed in
Section~\ref{sec:perf_model} of our present paper are kept to a
minimalistic style, in that we only rely on four easily
benchmarked/obtained hardware characteristic
parameters: $W_{\mathrm{thread}}^{\mathrm{private}}$,
$W_{\mathrm{node}}^{\mathrm{remote}}$,  $\tau$ and the last-level
cacheline size. This stands as a strong contrast to complicated
performance modeling approaches such as in~\cite{Li2015}.

%Related work on performance modeling of parallel SpMV implementations
%\& irregular inter-process communication (if any)

%% file: Sec_Heat2D.tex
\section{Performance modeling for a 2D uniform-mesh computation}
\label{sec:heat2D}

Our performance modeling strategy in Section~\ref{sec:perf_model}
was originally derived for the case of fine-grained irregular communication
that is due to indirectly indexed accesses to shared arrays. In this section, we will show that the same methodology, as well as the same hardware characteristic parameters, are applicable to other scenarios. As a concrete example, we will use an {\em existing} UPC implementation that solves a 2D heat diffusion equation $\frac{\partial \phi}{\partial t}=\frac{\partial^2\phi}{\partial x^2}+\frac{\partial^2\phi}{\partial y^2}$
on a uniform mesh. The UPC code was kindly provided by Dr.~Rolf Rabenseifner at HLRS, in connection with a short course on PGAS programming~\cite{PGAS_course2015}.

\subsection{Brief description of the code}
\label{sec:2Dcode}

\subsubsection{Data structure}
The global 2D solution domain is rectangular, so the UPC threads are arranged as a 2D processing grid, with {\tt mprocs} rows and {\tt nprocs} columns. (Note {\tt THREADS} equals {\tt mprocs*nprocs}.)
Each thread is thus identified by an index pair
{\tt (iproc,kproc)}, where
{\tt iproc=MYTHREAD/nprocs} and {\tt kproc=  MYTHREAD\%nprocs}.
The global 2D domain, of dimension $M\times N$, is evenly divided
among the threads. Each thread is responsible for a 2D subdomain of
 dimension $m\times n$, which includes a surrounding halo layer needed for communication with the neighboring threads. The main data structure consists of the following components:
\begin{quote}
\begin{verbatim}
shared [] double * shared xphi[THREADS];
double *phin;

xphi[MYTHREAD] = (shared [] double *) upc_alloc(m*n*sizeof(double));
phin= (double *)malloc(m*n*sizeof(double));
\end{verbatim}
\end{quote}
Note that the values in each shared array {\tt xphi[MYTHREAD]} have affinity to the allocating thread, but are accessible by all the other threads. Each thread also allocates a private array named {\tt phin} to store its portion of the numerical solution for a new time level, to be computed based on the numerical solution for the previous time level, which is assumed to reside in {\tt xphi[MYTHREAD]}. 

\subsubsection{Halo data exchange}
The halo data exchange, which is needed between the neighboring threads, is realized by that every thread calls {\tt upc\_memget} on each of the four sides (if a neighboring thread exists). In the vertical direction, the values to be transfered from the upper and lower neighbors already lie contiguously in the memory of the owner threads. There is thus no need to explicitly pack the messages. In the horizontal direction, however, message packing is needed before {\tt upc\_memget} can be invoked towards the left and right neighbors.
The following additional data structure is needed with packing and unpacking the horizontal messages:
\begin{quote}
\begin{verbatim}
/* scratch arrays for UPC halo exchange of non-contiguous data */
shared [] double * shared xphivec_coord1first[THREADS];
shared [] double * shared xphivec_coord1last [THREADS];
double *halovec_coord1first, *halovec_coord1last;

xphivec_coord1first[MYTHREAD] =
  (shared [] double *) upc_alloc((m-2)*sizeof(double)); 
xphivec_coord1last[MYTHREAD]  =
  (shared [] double *) upc_alloc((m-2)*sizeof(double));
halovec_coord1first = (double *) malloc((m-2)*sizeof(double)); 
halovec_coord1last  = (double *) malloc((m-2)*sizeof(double));
\end{verbatim}
\end{quote}

Consequently, the function for executing the halo data exchange is implemented as follows:
\begin{lstlisting}[caption=The halo data exchange function of an existing 2D heat equation solver,label=code_2D_comm]
#define idx(i,k) ((i)*n+(k)) 
#define rank(ip,kp) ((ip)*nprocs+(kp)) 

void halo_exchange_intrinsic()
{
  double* phi = (double *) xphi[MYTHREAD];
  int i,k;

  /* packing messages for the horizontal direction */
  if (kproc>0) {
    double *phivec_coord1first = (double *) xphivec_coord1first[MYTHREAD];
    for (i=0; i<m-2; i++)
      phivec_coord1first[i] = phi[idx(i+1,1)];
  }
  if (kproc<nprocs-1) {
    double *phivec_coord1last = (double *) xphivec_coord1last[MYTHREAD];
    for (i=0; i<m-2; i++)
      phivec_coord1last[i] = phi[idx(i+1,n-2)];
  }

  upc_barrier;

  /* message transfer and unpacking (needed for the horizontal direction) */
  if (kproc>0) {
    upc_memget(halovec_coord1first, xphivec_coord1last[rank(iproc,kproc-1)], (m-2)*sizeof(double));
    for (i=1; i<m-1; i++)
      phi[idx(i,0)] = halovec_coord1first[i-1];
  }
  if (kproc<nprocs-1) {
    upc_memget(halovec_coord1last, xphivec_coord1first[rank(iproc,kproc+1)], (m-2)*sizeof(double));
    for (i=1; i<m-1; i++)
      phi[idx(i,n-1)] = halovec_coord1last[i-1];
  }

  if (iproc>0)
    upc_memget(&phi[idx(0,1)], &(xphi[rank(iproc-1,kproc)][idx(m-2,1)]), (n-2)*sizeof(double));
  if (iproc<mprocs-1)
    upc_memget(&phi[idx(m-1,1)], &(xphi[rank(iproc+1,kproc)][idx(1,1)]), (n-2)*sizeof(double));
}
\end{lstlisting}

\subsubsection{Computation}

The computation that is carried out at each time level can be seen
in the following time loop:

\begin{lstlisting}[caption=The computational kernel of an existing 2D heat equation solver,label=code_2D_comp]
#define idx(i,k) ((i)*n+(k)) 
 
for (it=1; it<=itmax; it++) {
  double *phi = (double *) xphi[MYTHREAD];
  int i,k;

  /* communication per time step */
  halo_exchange_intrinsic();

  /* computation per time step */
  for (i=1; i<m-1; i++)
    for (k=1; k<n-1; k++)
      phin[idx(i,k)]=((phi[idx(i+1,k)]+phi[idx(i-1,k)]-2.*phi[idx(i,k)])*dy2i
                     +(phi[idx(i,k+1)]+phi[idx(i,k-1)]-2.*phi[idx(i,k)])*dx2i)*dt;

  /* copying the content of phin to phi, checking convergence etc. */
  // ....
}
\end{lstlisting}

\subsection{Performance modeling}

By slightly changing the formulas in Section~\ref{sec:5.2.5}, we can model
the cost of the different parts involved  
in the function {\tt halo\_exchange\_intrinsic} (Listing~\ref{code_2D_comm}) as follows:
\begin{eqnarray}
T^{\mathrm{halo,pack}}_{\mathrm{thread}} =T^{\mathrm{halo,unpack}}_{\mathrm{thread}}&=&
\frac{(S_{\mathrm{thread}}^{\mathrm{local,horiz}}
 +S_{\mathrm{thread}}^{\mathrm{remote,horiz}})(\mbox{\tt
    sizeof(double)} +\mbox{\tt sizeof}(\mathrm{cache\,line}))}
{W^{\mathrm{private}}_{\mathrm{thread}}}
\label{eq:halo_pack}\\
T^{\mathrm{halo,memget}}_{\mathrm{node}} &=&
\max_{\forall\,\mathrm{threads\;in\;node}}
\frac{2\cdot S_{\mathrm{thread}}^{\mathrm{local}}\cdot \mbox{\tt
    sizeof(double)}}{W^{\mathrm{private}}_{\mathrm{thread}}} \nonumber\\
&+&\sum_{\forall\,\mathrm{threads\;in\;node}} \left(C_{\mathrm{thread}}^{\mathrm{remote}}\cdot\tau
+\frac{S_{\mathrm{thread}}^{\mathrm{remote}}\cdot \mbox{\tt
    sizeof(double)}}{W^{\mathrm{remote}}_{\mathrm{node}}}\right) \label{eq:halo_memget}
\end{eqnarray}

Comparing (\ref{eq:halo_pack}) with (\ref{eq:pack_cost}) from Section~\ref{sec:5.2.5},
we can see that the cost due to indirect array indexing ({\tt sizeof(int)}) is no longer applicable.
We have also assumed in (\ref{eq:halo_pack}) that reading/writing from/to non-contiguous locations in the array {\tt phi} each costs a cache line. The value of $S_{\mathrm{thread}}^{\mathrm{local,horiz}}$ in (\ref{eq:halo_pack}) denotes the total volume of local and remote messages, per thread, to be transfered in the horizontal direction.
This can be precisely calculated when the values of $m$ and $n$, as well as the thread grid layout, are known.
Formula (\ref{eq:halo_memget}) is essentially the same as (\ref{eq:memput_cost_node})
from Section~\ref{sec:5.2.5}. It is commented that $S_{\mathrm{thread}}^{\mathrm{local}}$
in (\ref{eq:halo_memget}) denotes the total volume of all local messages (in both horizontal and vertical directions) per thread, likewise denotes $S_{\mathrm{thread}}^{\mathrm{remote}}$ the total volume of all remote messages, with $C_{\mathrm{thread}}^{\mathrm{remote}}$ denoting the number of remote messages per thread.
Putting them together, the total time spent on {\tt halo\_exchange\_intrinsic} is modeled as
\begin{eqnarray}
T^{\mathrm{halo}}_{\mathrm{2D}}=\max_{\forall\mathrm{nodes}}\left(
\left(\max_{\forall\,\mathrm{threads\;in\;node}}T^{\mathrm{halo,pack}}_{\mathrm{thread}}\right)
+T^{\mathrm{halo,memget}}_{\mathrm{node}}+
\left(\max_{\forall\,\mathrm{threads\;in\;node}}T^{\mathrm{halo,unpack}}_{\mathrm{thread}}\right)
\right).
\label{eq:halo}
\end{eqnarray}

The time spent on computation during each time step (see Listing~\ref{code_2D_comp})
is modeled as
\begin{eqnarray}
T^{\mathrm{comp}}_{\mathrm{2D}}=\frac{3(m-2)(n-2)\cdot\mbox{\tt sizeof(double)}}{W^{\mathrm{private}}_{\mathrm{thread}}}.
\label{eq:2Dheat_comp}
\end{eqnarray}
Here, we have assumed that every thread has exactly the same amount of
computational work. The value of $T^{\mathrm{comp}}_{\mathrm{2D}}$ is estimated on the basis of the minimum amount of traffic between the memory and the last-level cache, see e.g.~\cite{stengel2015quantifying}.

Table~\ref{tab:2D} shows a comparison between the actual time usages of
the 2D UPC solver with the predicted values of $T^{\mathrm{halo}}_{\mathrm{2D}}$ and
$T^{\mathrm{comp}}_{\mathrm{2D}}$. We have used the same hardware characteristic parameters as in Table~\ref{tab:comparison}. It can be seen that the predictions of
$T^{\mathrm{comp}}_{\mathrm{2D}}$ agree excellently with the actual measurements, 
while the prediction accuracy of
$T^{\mathrm{halo}}_{\mathrm{2D}}$ is on average 72\%.

\begin{table}[htb]
\caption{Comparison of actual and predicted time usages when
 running 1000 time steps of the 2D heat equation solver. The same hardware characteristic parameters as in Table~\ref{tab:comparison} are used.}
\label{tab:2D}
\centerline{%
\begin{tabular}{c|c||c|c||c|c}
\hline
\multicolumn{6}{c}{Mesh size $20000\times20000$}\\
\hline
{\small\tt THREADS}&{\small Partitioning}
&$T^{\mathrm{halo}}_{\mathrm{2D,actual}}$
&$T^{\mathrm{halo}}_{\mathrm{2D,predicted}}$
&$T^{\mathrm{comp}}_{\mathrm{2D,actual}}$
&$T^{\mathrm{comp}}_{\mathrm{2D,predicted}}$\\
\hline
16&$4\times4$&0.52&0.33&122.53&122.07\\
32&$4\times8$&0.44&0.37&61.55&61.04\\
64&$8\times8$&0.27&0.21&30.78&30.52\\
128&$8\times16$&0.29&0.21&15.31&15.26\\
256&$16\times16$&0.18&0.13&7.70&7.63\\
512&$16\times32$&0.14&0.14&3.85&3.81\\
\hline
\multicolumn{6}{c}{Mesh size $40000\times40000$}\\
\hline
{\small\tt THREADS}&{\small Partitioning}
&$T^{\mathrm{halo}}_{\mathrm{2D,actual}}$
&$T^{\mathrm{halo}}_{\mathrm{2D,predicted}}$
&$T^{\mathrm{comp}}_{\mathrm{2D,actual}}$
&$T^{\mathrm{comp}}_{\mathrm{2D,predicted}}$\\
\hline
16&$4\times4$&1.55&0.65&489.96&488.28\\
32&$4\times8$&1.08&0.73&246.25&244.14\\
64&$8\times8$&0.64&0.42&122.82&122.07\\
128&$8\times16$&0.64&0.42&61.85&61.04\\
256&$16\times16$&0.42&0.26&31.01&30.52\\
512&$16\times32$&0.29&0.26&15.47&15.26\\
\hline
\end{tabular}
}
\end{table}

%% file: Sec_Conclusion.tex
\section{Conclusion}
\label{sec:conclusion}

Our starting point is a naive UPC implementation of the SpMV $y=Mx$ in
Section~\ref{sec:3.2}. This naive implementation
excessively uses shared arrays, pointers-to-shared and {\tt upc\_forall}.
We have developed three increasingly aggressive
code transformations in Section~\ref{sec:transforms} aiming at
performance enhancement. The transformations include explicit thread
privatization that avoids {\tt upc\_forall} and casts
pointers-to-shared to pointers-to-local whenever possible, as well as
removing fine-grained irregular communications that are implicitly caused by
indirectly indexed accesses to the shared array {\tt x}. The latter
transformation is realized
by letting each thread adopt a private {\tt mythread\_x\_copy} array that
is prepared with explicit one-sided communications (using two
different strategies) prior to the SpMV computation.
Numerical experiments of a realistic application of SpMV and the associated
time measurements reported in Section~\ref{sec:experiments} have
demonstrated the performance benefits due to the code transformations.
The performance benefits are also justified and quantified by the three
performance models proposed in Section~\ref{sec:perf_model}.

While the code transformations lead to improved performance,
the complexity of UPC programming is increased at the same time.
(Trading programmability for performance is by no means specific
for our special SpMV example. The textbook of UPC~\cite{UPC_book} has
ample examples.)
The naive UPC implementation in Section~\ref{sec:3.2}
shows the easy programmability of UPC that is fully comparable with OpenMP, as discussed
in e.g.~\cite{Marowka2005}.
The first code transformation, in form of explicit thread
privatization shown in Section~\ref{sec:4.1}, may be done by
automated code translation. The second and third code
transformations, see Sections~\ref{sec:4.2}-\ref{sec:4.3},
are however more involved. The adoption of one {\tt mythread\_x\_copy} array
per thread also increases the memory footprint.
Despite reduced programmability, all the UPC implementations
maintain some advantages over OpenMP in targeting
 distributed-shared memory systems and promoting data locality.
The third code transformation, UPCv3, results in a programming style quite similar
to that of MPI.
Nevertheless, UPCv3 is easier to code
than MPI, because global indices are retained for accessing the entries of array {\tt
  mythread\_x\_copy}.
An MPI counterpart, where all arrays are explicitly partitioned among
processes, will have to map the global indices to local indices. Moreover, 
one-sided explicit communications via UPC {\tt upc\_memget} and {\tt
  upc\_memput} functions are easier to use. Performance advantage of
UPC's one-sided communication over the MPI counterpart has also been
reported in e.g.~\cite{ibrahim2014evaluation}.
On the other hand, persistent advantages of MPI over UPC include better
data locality and more flexible data partitionings. A comparison of
performance and programmability between UPC and MPI was given
in~\cite{Prugger2016} for a realistic fluid dynamic implementation.
For a general
comparison between OpenMP, UPC and MPI programming, we refer to~\cite{Shan2010}.

It should be stressed that the SpMV computation is chosen for this
paper as an illustrating example of fine-grained irregular
communication that may arise in connection with naive UPC
programming. The focus is not on the SpMV itself, but on the
code transformations and the performance models in general.
Moreover, our performance models are to a great extent independent of the
UPC programming details, but rather focusing on the incurred
communication style, volume and frequency. The hardware characteristic
parameters $W_{\mathrm{thread}}^{\mathrm{private}}$,
$W_{\mathrm{node}}^{\mathrm{remote}}$ and the cacheline size are
equally applicable to similar communications and memory-bandwidth
bound computations implemented by other programming models than UPC.
Even the latency of individual remote memory accesses, $\tau$, can
alternatively be measured by a standard MPI ping-pong benchmark.
Our philosophy is to
represent a target hardware system by only four
characteristic parameters, whereas the accuracy of the performance
prediction relies on an accurate counting of the incurred communication
volumes and frequencies. Accurate counting is essential and thus
cannot be generalized, because
different combinations of the problem size, number of threads, and block
size will almost certainly lead to different levels of performance for the same
parallel implementation.